%% file: main.tex
\documentclass[fleqn,usenatbib,useAMS]{mnras}
\usepackage{graphicx}
\usepackage{amsmath}
\usepackage{multicol}
\usepackage{bm}
\usepackage{pdflscape}	
\usepackage{multirow}
\usepackage[dvipsnames]{xcolor}
\definecolor{darkteal}{RGB}{0, 89, 94}
\usepackage{makecell}
\usepackage{array}
\usepackage{color}
\usepackage{boldline}
\usepackage{subcaption}

\input{macros.tex}

\usepackage{orcidlink}
\newcommand{\orcid}[1]{\,\orcidlink{#1}}

\usepackage[T1]{fontenc}
\usepackage{ae,aecompl}
\usepackage{newtxtext,newtxmath}

\title[No parametrisation, No Problem]{No parametrisation, No Problem: A Weakly Modelled Framework to Constrain the Luminosity Distance--Redshift Relation Using Gravitational Wave Sirens}
\author[E. Colangeli et al.]{%
Elena Colangeli$^{1}$\orcid{0009-0009-9783-3407}\thanks{E-mail: elena.colangeli@port.ac.uk},
Konstantin Leyde$^{2}$\orcid{0000-0001-7661-2810},
Tessa Baker$^{1}$\orcid{0000-0001-5470-7616}
and Anson Chen$^{3}$\orcid{0000-0001-9174-7780}
\\
$^{1}$Center for Computational Astrophysics, Flatiron Institute, 162 5th Ave, New York, NY 10010\\
$^{2}$Institute of Cosmology and Gravitation, University of Portsmouth, Burnaby Road, Portsmouth PO1 3FX, United Kingdom\\
$^{3}$International Centre for Theoretical Physics Asia-Pacific, University of Chinese Academy of Sciences, Beijing, 100190, China
}

\begin{document}
\label{firstpage}
\pagerange{\pageref{firstpage}--\pageref{lastpage}}
\maketitle

\begin{abstract}
Cosmological tests of general relativity (GR) using gravitational waves (GWs) often rely on parametrised forms of the luminosity distance-redshift ($ d_{\rm L} - z$) relation, which is modified in alternative theories of gravity where the gravitational coupling strength is time-dependent. Although they can lead to stringent results, these parametrisations encapsulate a restricted range of behaviours that may not represent the full variety of modified gravity theories. 
We present a quasi-model-independent framework to reconstruct deviations from GR in the ratio of the GW to standard electromagnetic luminosity distance. We find that our method can confidently constrain monotonic, irregular, and oscillatory features in the ratio, without biasing population distribution parameters. Applying our framework to GWTC-5 data, we find no deviations from GR.
We report that O5-like GW data will not resolve deviations from GR smaller than 30\% with our method; however, third generation GW detectors can confidently detect deviations of order 5\% to 3$\sigma$ significance. 
\end{abstract}

\section{Introduction}
\label{sec:intro}
The recent increasing number of gravitational wave (GW) events has consolidated GW cosmology as a complementary cosmological probe to electromagnetic (EM) results \citep{LIGOScientific:2026uyd, LIGOScientific:2025jau}. 
In addition to constraining parameters of the $\Lambda$CDM model, most notably the Hubble constant ($H_0$), it is possible to use GW data to test deviations from general relativity (GR) on cosmological scales\footnote{Tests of strong field deviations can also be performed, using waveforms, dispersion measures, ringdown and other analyses \citep{LIGOScientific:2026qni}.}.

A leading aspect of this is to constrain the relation between luminosity distance $d_{\rm L}$ and redshift $z$. When working with bright sirens, or gravitational wave events with an electromagnetic (EM) counterpart, one can directly compare luminosity distances with the measured redshift of the host galaxy. In order to obtain conclusive constraints from bright sirens, many events are required \citep{Nishizawa:2019rra, DAgostino:2019hvh, Belgacem:2019tbw, Colangeli:2025bnb}. As only one such event has been confirmed \citep{LIGOScientific:2017zic}, we instead turn to dark sirens (events without EM counterparts) which the LIGO-Virgo-KAGRA (LVK) collaboration has observed in the hundreds \citep{LIGOScientific:2026wfs}, and which are already being used to perform cosmological tests of GR \citep{LIGOScientific:2021aug, LIGOScientific:2025jau, LIGOScientific:2026uyd}.
Since redshifts cannot be measured directly for this type of event, several frameworks have been developed to break the inherent redshift-distance degeneracy. Methods to do so include leveraging population properties of the GW sources, a technique known as `spectral sirens' \citep{Mastrogiovanni:2021wsd, Mancarella:2021ecn, Ezquiaga:2022zkx, Mastrogiovanni:2023zbw, Bertheas:2026odj}, constructing a line-of-sight redshift prior in the localisation area of each signal derived from the galaxy catalogue \citep{DelPozzo:2011vcw, Gray:2019ksv}, or cross-correlating the angular distribution of GW events to large scale structure maps from galaxy surveys \citep{Oguri:2016dgk, Mukherjee:2018ebj, Mukherjee:2020hyn, Ferri:2024amc, SantiagodeMatos:2025iyj}.
In this work we follow the spectral sirens method, which we will in detail in Section~\ref{ssec:spectral sirens} only using the properties of the population mass distribution of binary black holes (BBHs)\footnote{Note that the term `spectral sirens' refers to analyses using mass distribution only; on the other hand, the `dark sirens' method refers to simultaneously using the spectral sirens method and galaxy catalog information. This is in like with LVK definitions.}. 

Many works in the literature \citep{Belgacem:2017ihm, Mukherjee:2020mha, Finke:2021aom, Leyde:2022orh, Chen:2023wpj, Colangeli:2025bnb, Chen:2026owi, Chen:2026htz} including the latest LVK collaboration cosmology results \cite{LIGOScientific:2026uyd, LIGOScientific:2025jau} have constrained modified gravity (MG) effects on the $d_{\rm L} - z$ relation using parametrised forms (see section VI of \cite{Ezquiaga:2018btd} for an overview of some popular ones). 

The most common parametrisations derive from the $\Xi_0-n$ ansatz (from \cite{Belgacem:2018lbp}) and the $\alpha_{\rm M}$ description of Horndeski theory \citep{Bellini:2014fua}, where the standard is to adopt phenomenological forms (e.g. $\alpha_{\rm M} \propto c_{\rm M}\Omega_\Lambda$).
\footnote{These are the parametrisations used in the aforementioned LVK cosmological results} 
These parametrisations yield monotonic functions, regulated by one (or two, in the case of $\Xi_0-n$) parameters.
Though practical, as they only require a small number of free parameters, these functional forms fail to accurately capture the behaviour of the full space of MG theories. For example, \cite{Linder:2016wqw} explains how certain parametrisations can fail, and \cite{Ezquiaga:2021ler, Ezquiaga:2018btd, LISACosmologyWorkingGroup:2019mwx} show a slew of theories with non-trivial $d_{\rm L} - z$ relations.

To move beyond such restrictions this work introduces a quasi-model-independent, or `weakly modelled', method for constraining the $ d_{\rm L} - z$ relation. By `quasi' model-independent we mean that despite not choosing a fixed functional form governing the the modified $ d_{\rm L} - z$ relation, there still are a set number of parameters that determine its properties. However, our method is able to access a much wider array of functional forms for the deviations from GR predictions, which are not motivated by a specific MG theory. A few other methods for non-parametric reconstructions of the luminosity distance-redshift relation have been proposed \citep{Belgacem:2019zzu,Afroz:2024oui, Afroz:2023ndy,Vallejo-Pena:2026tmy}, relying on techniques such Gaussian Processes, scaling anchored by EM standard rulers (BAO) or candles (supernovae), and piecewise interpolations. We present a  complementary method implemented in a new pipeline, \texttt{CosmoPyro} \citep{COSMOPYRO}, where we model the $d_{\rm L} - z$ relation via a Fourier expansion. Using this weakly-modelled relation and the mass distribution method (with a parametrised form in this case) we jointly constrain deviations from GR and source properties.

Efforts to steer away from parametric ansatzes are being made to also constrain the source mass frame distribution of black holes \citep{Tagliazucchi:2026gxn, LIGOScientific:2025pvj, LIGOScientific:2026ctl, Gennari:2026dfy}. We do not explore this here as our focus is on opening up the accessible space of MG constraints; we refer the reader to \cite{COSMOPYRO} for implementations in \texttt{CosmoPyro}. 
Note also that we will refrain from using the dark siren galaxy catalog method at this point. Galaxy catalog information improves cosmological results at $2-8\%$, mainly due to its incompleteness at intermediate to high redshifts \citep{LIGOScientific:2025jau, LIGOScientific:2026uyd}, where the majority of events lie. We will additionally provide a forecast for future bright sirens data to showcase the full capabilities of this new method. Finally, we will apply our methodology to real events, using dark siren data from the latest LVK data release \citep{LIGOScientific:2026wfs}.

This paper is structured as follows: in Section~\ref{sec:theory} we introduce the theoretical background of modified GW propagation and spectral sirens cosmology; in Section~\ref{sec:statistical framework} we present the statistical framework of our analyses. In Section~\ref{sec:method} we give an overview of the specific tools we use to constrain modified gravity, presenting our findings in Section~\ref{sec:results}. We conclude with a summary and final remarks in Section~\ref{sec:conclusions}.

\section{THEORY}
\label{sec:theory}
\subsection{Modified GW distance}
\label{ssec:modified gw distance}
We carry out a joint analysis probing population characteristics of the mergers and modified gravity effects.  
We fix the background cosmology to $\Lambda$CDM throughout to isolate the effects of modified gravity on the GW distance. This choice is motivated by the strong constraints on background parameters from EM probes (e.g.  \cite{Planck:2018vyg, Brout:2022vxf})\footnote{While the Hubble tension presents an ongoing challenge for the local value of the expansion rate $H_0$, we explicitly address this later by fixing its value to to both early and late-time results.}.
For a given redshift, the luminosity distance is defined as:  
\begin{align}
    d_{\rm L} = \frac{(1+z)}{H_0} \int_0^z \frac{ \dd z'}{E(z')} \text{ ,}
\end{align}
where $E(z) = H_0^{-1}(\sqrt{\Omega_r(1+z)^4+\Omega_m(1+z)^3+\Omega_k(1+z)^2+\Omega_\Lambda})$, with the $\Omega$ terms corresponding to radiation, matter, curvature, and dark energy densities in the universe. Our background is fixed to $\Lambda$CDM with $\Omega_m = 1-\Omega_\Lambda = 0.3$. 
In scalar/vector/tensor-tensor theories the measured distance from the GW signal can differ from this form of $d_{\rm L}$ via a non-minimal coupling of new fields to the Ricci scalar. When computing the propagation equation of gravitational waves, the coupling's derivative can produce a non-zero friction term \citep{Saltas:2014dha, Amendola:2017ovw, Belgacem:2017ihm, Belgacem:2018lbp, Ezquiaga:2021ler, Romano:2022jeh}. This friction term explicitly carries into the amplitude of the signal, which encodes the distance information.

This GW and EM distances difference causes their ratio to deviate from 1 (the GR result). Here we aim to constrain this ratio as a function of redshift:
\begin{align}
    r(z) = \frac{\dgw(z)}{d_{\rm EM}(z)} \, .
\end{align}
In this work $d_{\rm EM}$ is specifically the EM distance, while we maintain the more general nomenclature $d_{\rm L}$ for luminosity distance of either domain.

\subsection{Spectral Sirens}
\label{ssec:spectral sirens}
Gravitational wave signals encode information about the detector-frame masses of the merging bodies, which are redshifted from the source-frame: $m_{\rm det} = (1+z)m_{\rm s}$. 
One can break the degeneracy between redshift and mass measurement if the source-frame mass distribution is known: measuring the shift in detector-frame mass and comparing it to source-frame mass we can pick out the redshift. This is analogous to using spectra for galaxies, hence the name of `spectral sirens', though unlike absorption or emission lines we do not have much theoretical knowledge of where the analogous mass features lie. As a consequence, we perform the analysis on a population level (rather than on a per-event basis), where markers in the mass distribution of all sources such as peaks and underdensities shift based on distance and cosmology. 
This requires prior knowledge of the source-frame mass distribution. As we do not know the distribution with certainty, the analysis jointly constrains $r(z)$ and the parameters of the mass distribution. 
 
Currently-used models for the prior of this mass distribution, are the \textit{\textsc{PowerLaw+Peak} (PLP)}, \textit{\textsc{Multi Peak} (MLTP)} and \textit{\textsc{\textsc{FullPop}}} (used in the LVK cosmology results \cite{LIGOScientific:2026uyd}). These are `strongly modelled' parametric distributions, where the distribution's shape is fixed, with parameters controlling the strength, width and position of each feature. The first two only include black holes, and features of the mass spectrum (peaks, over/underdensities) are informed by astrophysics and observed GW events. In particular, the PLP model is a power-law slope with a gaussian peak, and the MLTP has a power law slope and two peaks. The \textsc{FullPop} model includes neutron stars as well as black holes on a continous mass spectrum. 
We only simulate black hole mergers and choose the MLTP model, as it is favoured by the data \citep{LIGOScientific:2026uyd, LIGOScientific:2026ctl} over PLP\footnote{We choose MLTP as we deal exclusively with BBHs, but note that \textsc{FullPop} is overall favoured by the data, as some events lie in the borderline region between the low-mass black holes and the heavy neutron stars.}. 
Furthermore, when analysing real data we will focus only on mergers where both objects are classified as black holes. 

This distribution is described by 10 parameters: 
\begin{itemize}
    \item $\alpha$: slope of the primary mass power law,
    \item $\delta_{\rm m}$: low mass tapering,
    \item $m_{\rm min}$: minimum primary mass,
    \item $m_{\rm max}$: maximum primary mass,
    \item $\mu_{\rm g, low}$: mean of the lower mass gaussian peak,
    \item $\mu_{\rm g, high}$: mean of the higher mass gaussian peak,
    \item $\sigma_{\rm g, low}$: standard deviation of the lower mass gaussian peak,
    \item $\sigma_{\rm g, high}$: standard deviation of the higher mass gaussian peak,
    \item $\lambda_{\rm g}$: fraction of sources in both gaussian peaks,
    \item $\lambda_{\rm g, low}$: fraction of sources in the lower mass peak.
\end{itemize}
We include these parameters in our analyses\footnote{We additionally infer the slope of the secondary mass distribution, which we introduce in Section~\ref{ssec:mock data}} along with the ratio parameters and the redshift distribution parameter $\gamma$. This parameter (formally defined later in Section~\ref{ssec:mock data}) sets the steepness of the power law in the redshift prior, effectively regulating the number of events at a given redshift, making it highly degenerate with modifications to GW distances. 
Note that for simplicity we do not explicitly sample over spins, as utilise an approximate model to generate our mock data (see Section~\ref{ssec:mock data}). Nevertheless, non-trivial spin distributions are favoured by the latest LVK cosmology analyses \citep{LIGOScientific:2026uyd}.

\section{Statistical framework}
\label{sec:statistical framework}
In this section we present the statistical framework we use to jointly infer $r(z)$ and GW population parameters. We employ hierarchical Bayesian inference, the standard framework for gravitational wave analyses \citep{Mandel:2018mve, Gray:2023wgj, Mastrogiovanni:2023zbw}. For observed GW dataset $x$, we can use Bayes' theorem to obtain
\begin{align}
    p({\Lambda}|x) = \frac{\mathcal{L}(x|{\Lambda})\pi({\Lambda})}{p(x)} \, ,
\end{align}
where ${\Lambda} = \left\{\Lpop, \Lm\right\}$: the first term contains the source frame mass population parameters we wish to infer (and additional redshift distribution parameter $\gamma$), 
while the second one represents the modified distance ratio; note that we fix cosmology, hence it does not appear in our derivation). The numerator contains the likelihood $\mathcal{L}(x|{\Lambda})$ of observing the data for given population and cosmological models, while the terms $\pi({\Lambda})$ and $p(x)$ are the priors on the model parameters to infer, and the evidence, respectively.

The hierarchical likelihood can be expressed as a product of each single event likelihood, since observations are independent of one another:
\begin{align}
\label{eq:hierarchical likelihood}
    \mathcal{L}(x|\Lpop, \Lm) \propto \prod_{i =1}^{N_{\rm obs}} \frac{\mathcal{L}(x_i|\Lpop, \Lm)}{\beta(\Lambda)}  \, ,
\end{align}
Our event data vector $x_i$ contains the detector frame primary mass of the binary $m^{\rm d}_{1,i}$, mass ratio $q_i$ (defined as the ratio between secondary and primary mass), and GW luminosity distance ${\dgw}_i$, which allows us to expand our likelihood
\begin{align}
\label{eq:event likelihood}
    \nonumber \mathcal{L}(x_i|&\Lpop, \Lm) \propto p(m^{\rm d}_{1,i}, q_i, {\dgw}_i | \Lpop, \Lm)\\ \nonumber
    = &\iint p(m^{\rm d}_{1,i}, q_i, {\dgw}_i, m^{\rm s}_{1,i}, z_i |\Lpop, \Lm) \dd m^{\rm s}_{1,i} \dd z \\ \nonumber
    = &\iint p(m^{\rm d}_{1,i}, {\dgw}_i | m^{\rm s}_{1,i}, q_i, z_i, \Lm) p(m^{\rm s}_{1,i}, q_i | \Lpop) \\ & p(z_i | \Lm) \dd m^{\rm s}_{1,i} \dd z \,.
\end{align}
In the second step we have introduced $m^{\rm s}_{1,i}$, the primary mass of the event in source frame, and $z$, the redshift. We can simplify the expression using the relation between detector and source frame masses $m^{\rm d}_{1,i} = m^{\rm s}_{1,i} (1+z_i)$  
We additionally rewrite the GW luminosity distance as a function of redshift and cosmology to obtain 
\begin{align}
    \nonumber p(m^{\rm d}_{1,i}, {\dgw}_i |& m^{\rm s}_{1,i}, q_i, z_i, \Lm) \\ \nonumber = &\delta(m^{\rm d}_{1,i} -\hat{m}^{\rm d}_{1,i}(m^{\rm s}_{1,i}, q_i, z_i))\delta(d_{\rm GW, i} - \hat{d}_{\rm GW}(z_i, \Lm))
    \\ \nonumber = &\delta(m^{\rm s}_{1,i} - \hat{m}^{\rm s}_i(m^{\rm d}_{1,i},  q_i, z_i)) \times \\ &\times \delta(z_i - \hat{z}_i({\dgw}_i, \Lm))\left|\frac{\partial(m^{\rm d}_{1,i},{\dgw}_i)}{\partial(m^{\rm s}_{1,i}, z_i)}\right|
\end{align}
which collapse the single event likelihood in Eq.~\eqref{eq:event likelihood} into
\begin{align}
    \nonumber  \mathcal{L}(m^{\rm d}_{1,i}, {\dgw}_i &| m^{\rm s}_{1,i}, q_i, z_i, \Lm) \\&= p(\hat{m}^{\rm s}_i, q_i|\Lpop)p(\hat{z}_i|\Lm)\left|\frac{\partial(m^{\rm d}_{1,i},{\dgw}_i)}{\partial(m^{\rm s}_{1,i}, z_i)}\right| \, .
\end{align}
In Eq.\eqref{eq:hierarchical likelihood} we also introduce the term $\beta(\Lambda)$ which represents the fraction of all detectable events for a given model $\Lambda$, to correct for selection bias in GW observations. This is defined as $\beta(\Lambda) = \int{P_{\rm det}(\theta)p_{\rm pop}(\theta|\Lambda)}\dd\theta$. The detection probability $p_{\rm det}$ is integrated over the parameters $\theta$ of the binaries. 
Following standard cosmological analyses \citep{Mandel:2018mve}, we compute it using a Monte Carlo approximation:
\begin{align}
\label{eq:injections}
    \beta(\Lambda) \approx \frac{1}{N_{\rm inj}} \sum_{i=1}^{N_{\rm det}} \frac{p(m^{\rm s}_{1,i}, m^{\rm s}_{2,i}, z_i| \Lambda)}{\pi_{\rm draw}(m^{\rm s}_{1,i}, m^{\rm s}_{2,i}, z_i| \Lambda)} \, .
\end{align}
$N_{\rm det}$ is the fraction of total injections $N_{\rm inj}$ (simulated events) above the detection threshold, 
which should be scaled according to the number of detected events \citep{Heinzel:2025ogf}. We divide by the initial probability with which we draw the injection $\pi_{\rm draw}$. When drawing our injections it is crucial to ensure good coverage of all of parameter space, here represented in source frame masses $m^{\rm s}_{1}, m^{\rm s}_{2}$, and redshift $z$. For the mock data case the detection threshold is a signal-to-noise ratio (SNR) of 12, while for real data we employ the false alarm rate metric (FAR), and only include events with FAR$< 0.25~{\rm yr^{-1}}$, the same selection criterion used in the LVK cosmology analyses. 

Note that the bright sirens case, knowing the redshift of each event renders population parameters encoded in the $\Lpop$ term from Eq.~\ref{eq:hierarchical likelihood} negligible, as the inference becomes less sensitive to the underlying mass distribution. Hence we utilise a simplified single-event bright siren likelihood, dependent only on redshift:

\begin{align}
\label{eq:bright likelihood}
    \mathcal{L}_{\rm bright}(x_i|\Lm) \propto \int \delta(d_{\rm GW, i} - &\hat{d}_{\rm GW}(z_i, \Lm))p(z_i|\Lm)\dd z_i \,.
\end{align}

\section{Method}
\label{sec:method}
\subsection{Package}
\label{ssec:package}
In GW cosmology, dark sirens packages such as \texttt{gwcosmo} \citep{Gray:2019ksv,Gray:2021sew, Gray:2023wgj}, \texttt{icarogw} \citep{Mastrogiovanni:2023zbw}, and \texttt{CHIMERA} \citep{Borghi:2023opd, Tagliazucchi:2025ofb}, are inference frameworks designed to measure cosmological parameters implementing the spectral sirens and galaxy catalog methods. 
As previously mentioned, we utilise a new pipeline, \texttt{CosmoPyro}\footnote{\url{https://github.com/KonstantinLeyde/CosmoPyro}}\citep{COSMOPYRO}, which is a gradient-based dark siren code. The gradients allow for sampling of high-dimensional posteriors using the No-U-Turn Sampler (NUTS). This is a modification of the Hamiltonian Monte Carlo \citep{Duane:1987de} which has the advantage of automatically stopping a trajectory once it turns around \citep{hoffman2014no}. In order to compute these gradients, the pipeline is built using the \texttt{NumPyro} library \citep{phan2019composable, bingham2019pyro}. Leveraging GPUs we can perform modified gravity analyses of 500 events in 7 hours on average for 500 warmup steps and 1000 posterior samples. 

\subsection{Reconstruction of $r(z)$}
\label{ssec:reconstruction of r(z)}
We follow the method proposed in \cite{Leyde:2024tov} for galaxy map reconstructions and adapt it in order to reconstruct $r(z)$. 
This method utilises a Fourier reconstruction, which allows us to reproduce any type of function due to the completeness of this basis: as one increases the number of modes in the basis, the reconstruction converges towards the true functional form of the ratio. This ensures a higher flexibility than the parametrised methods, as we will demonstrate below.

We start by drawing $N$ independent samples from a Gaussian $\mathcal{N}(0, 1)$, which form a whitened random field in real space. We then transform them into Fourier space to obtain the Fourier modes $F(k)$ and reweight each mode by the square root of a power spectrum:
\begin{align}
\label{eq:F(k)}
    F(k)_{\rm weighted} = \sqrt{P(k)}F(k)\, . 
\end{align}
We choose the form of the power spectrum to be
\begin{align}
\label{eq:P(k)}
    P(k) = \frac{A}{\ell[1 + (\ell k)^{a}]}\, ,
\end{align}
where $k$ is the wavenumber, $A$ is the amplitude of the power spectrum and $\ell$ sets the correlation scale. The parameter $a$  is an arbitrary parameter which controls the general shape of $P(k)$ in order to concentrate power at low-$k$ (large scales, low frequencies), while still allowing for smaller, higher frequency oscillations. The value of $a$ is chosen (along with the correlation scale $\ell$, as we will see) 
to avoid fictitious small-scale structure, since we expect the scale of features in $r(z)$ to be of the same order of magnitude as the Hubble scale.

\begin{figure*}
    \makebox[\textwidth][c]{
        \begin{minipage}{1.1\textwidth}
            \centering
            \begin{subfigure}[b]{0.52\linewidth}
                \centering
                \includegraphics[width=\linewidth, trim={1cm 0cm 0cm 1cm}, clip]{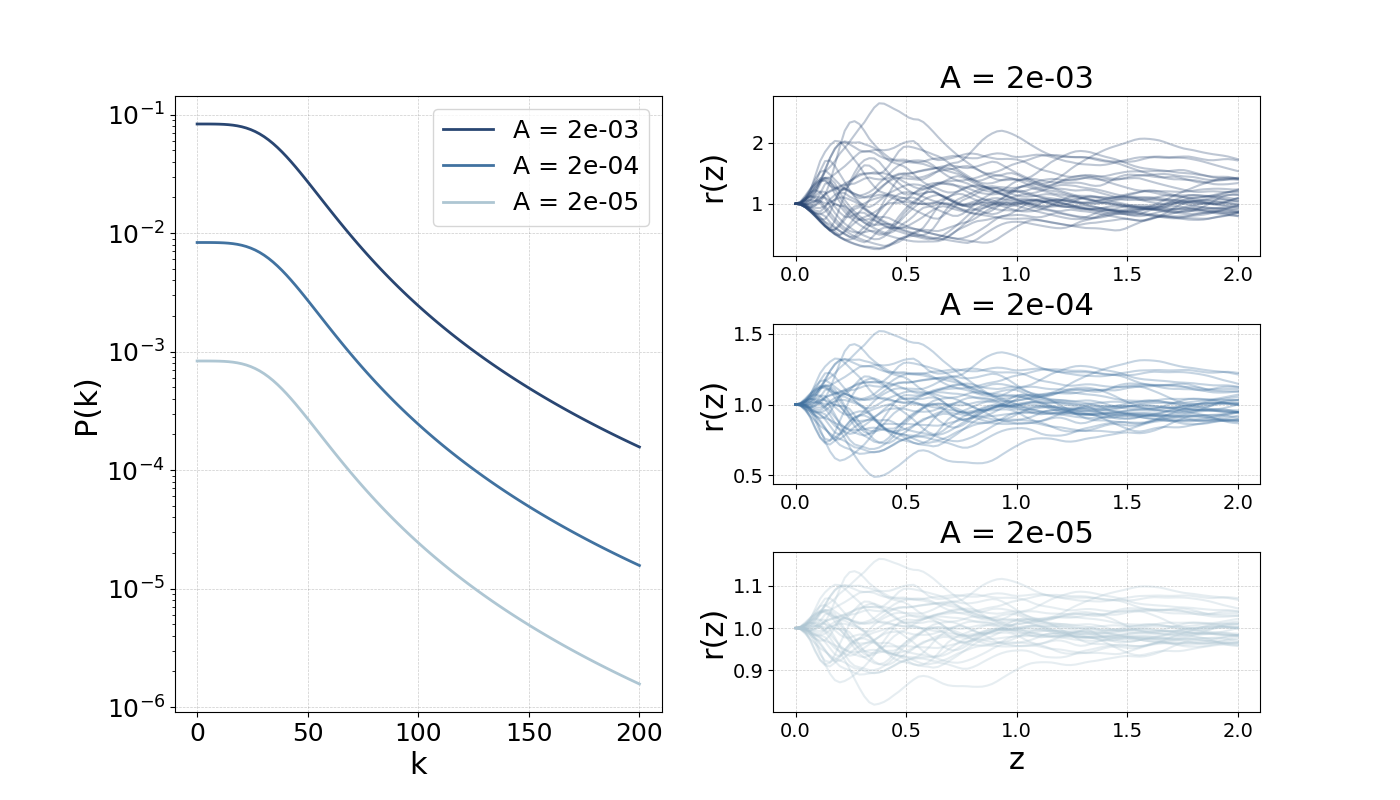}
                \caption{}
                \label{fig:amplitude}
            \end{subfigure}%
            \hspace{-1cm}
            \begin{subfigure}[b]{0.52\linewidth}
                \centering
                \includegraphics[width=\linewidth, trim={1cm 0cm 0cm 1cm}, clip]{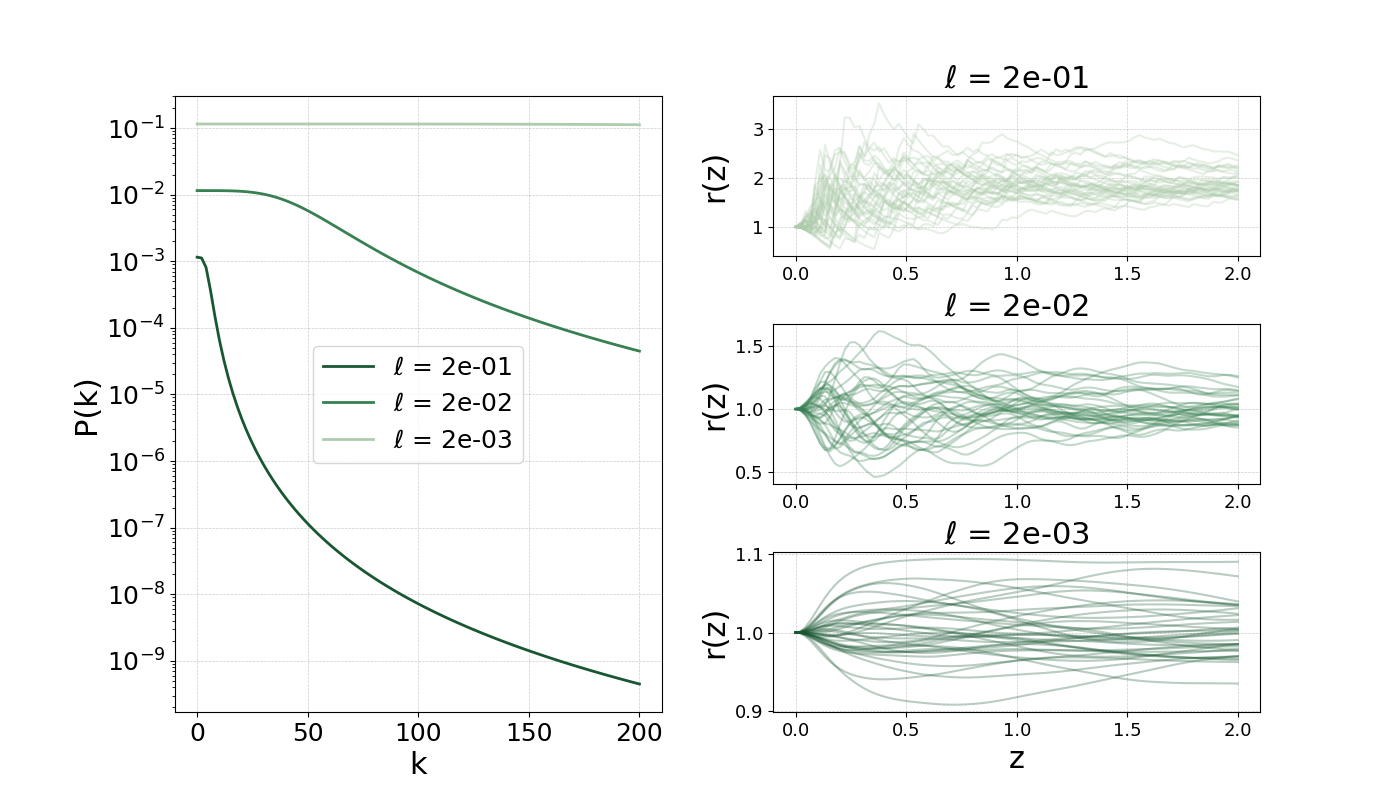}
                \caption{}
                \label{fig:correlation}
            \end{subfigure}
        \end{minipage}
    }
    \caption{Depiction of the impact that the power spectrum $P(k)$ has on the ratio $r(z)$, focusing on the impact of the amplitude $A$ (left) and correlation scale parameter $\ell$ (right). In both cases we depict power spectrum as a function of wavenumber on the left-hand-side panel, plotting three different orders of magnitude of $A$ (left) and $\ell$ (right). On the right panel we show the ratio corresponding to the three different values. When varying $A$ on the left we fix $\ell=0.02$; when varying and $\ell$ on the right we fix $A=0.0002$. In both cases darker colour corresponds to a larger number for the parameter. All realisations satisfy the constraint that $r(z)\to 1$ as $z\to0$. 
    }
    \label{fig:P(k)}
\end{figure*}
A visual representation of how amplitude and correlation influence our reconstructions is presented in Figure~\ref{fig:P(k)}. In Figure~\ref{fig:amplitude} we quantify the change in ratio amplitude caused by a change in power spectrum amplitude. We plot $P(k)$ for three orders of magnitude of $A$, where a darker colour indicates a larger amplitude. An increase in amplitude of an order of magnitude corresponds to a $\sqrt{10}$ increase in the amplitude of the ratio. We show the analogous plot for three orders of magnitude of the correlation scale $\ell$ in Figure~\ref{fig:correlation} (again darker colour corresponds to higher $\ell$). As $P(k) \propto 1/\ell$ approximately, a larger value corresponds to more correlation power at large scales. We indeed see the features in the ratio getting progressively smoother as $\ell$ grows. While the available modes are fixed by the initial sample draw, the data informs the features of the corresponding oscillatory functions (amplitude and phases of each mode). 

We apply two additional physical constraints. Firstly, we expect any deviation from GR to vanish at small redshifts to satisfy local constraints. In order to do so we introduce a suppression factor of $1 - \exp(-z/z_*)^2$ into $r(z)$ which forces the ratio to tend to 1 as $z \to 0$. We infer $z_*$, the low redshift scale at which this transition occurs. Secondly, we want to avoid any turnover in GW distance, which can be caused by our oscillatory basis. We achieve this by computing $\dgw$ via a cumulative sum of its derivative in GR, which is the derivative of EM distance: $\dgw(z) \propto \sum_i [\dd d_{\rm EM}/\dd z]_i\Delta z_i$. By ensuring the derivative stays positive, we ensure $\dgw$ cannot turn downward. The tradeoff is that the priors in ratio becomes mildly asymmetric, preferring positive ratio values and cutting off unphysical regions. This is visible in Fig. \ref{fig:P(k)} were for select values of $A$ and $\ell$ we observe that the prior draws favour $r(z)>1$.
It is important however to remember that though both GW and EM distances are monotonic by definition, their ratio need not be. 
To summarise, the weakly-modelled definition of the ratio, evaluated at redshift nodes $z_i$, is given by
\begin{align}
    r(z) =\frac{1}{d_{\rm EM}} \sum_{i=0}^n\left\{1+F(z_i)\times\left[1 - \exp\left(-\frac{z_i}{z_*}\right)^2\right]\right\}\times \left[\frac{\dd d_{\rm EM}}{\dd z}\right]_i\Delta z \, ,
\end{align}
where $F(z)$ is the inverse Fourier transform of $F(k)_{\rm weighted}$ from Eq.~\eqref{eq:F(k)} which multiplies the suppression term and the derivative term. The sum over modes runs up to $n$, which we fix in our runs, and we evaluate everything over redshift intervals $\Delta z$. Because higher frequency Fourier modes introduce finer structures in the reconstruction, we scale the number of redshift intervals with the number of basis functions to be able to resolve these. For $n$ modes we have $n$ redshift intervals, or effectively $n+1$ nodes in redshift at which we evaluate the ratio. Our analyses infer the phases and amplitudes of the Fourier modes, encoded in $F(z_i)$, and the the suppression parameter $z_*$. The power spectrum hyperparameters are fixed for now to the values we quote in the next section.

\subsection{Caveats}
\label{ssec:caveats}
Though Fourier decompositions can in principle describe any function, some caution has to be used when applying this method onto a physical problem. 

Firstly, one needs to keep in mind that, as shown in Figure~\ref{fig:P(k)}, amplitude and correlation scale of the power spectrum produce correlated effects onto the amplitude and correlation of the ratio itself. Hence, when choosing a correlation scale which can resolve features in the ratio, one has to choose the appropriate amplitude to probe the desired deviations from $1$. For our redshift range we pick power spectrum parameters of $A = 2.3 \times 10^{-4}$, $\ell = 2.4 \times 10^{-2}$, and $a = 4$, which can resolve fluctuations of order $\Delta z \sim 0.04$ and has an average deviation from $r(z) = 1$ of $\sim 20\%$ 
. Additionally, a sufficient number of basis functions should be chosen, and the number of redshift nodes at which the ratio is computed scales with this value. These choices need to be accurately made depending on the redshift range in question, a larger range would require more nodes, and hence more basis functions\footnote{Working in Fourier space allows for this number to be scaled up without having to greatly increase computing power.}. 

Another caveat has been mentioned above: our prior space can favour regions where $r(z)>1$. This only occurs when the amplitude is such that a lot of draws hit the negative $\dd \dgw / dz$ wall. It can be avoided by studying one's prior configurations before running any analysis, as this asymmetry normally arises for power spectrum settings that allow the ratio to reach unphysically large deviations from GR (see e.g. the top panel in Figure~\ref{fig:correlation}, where deviations of $r(z)$ reach 200\%). We also wish to point out that with sufficiently good and numerous data any ratio functional form \textit{can} be reconstructed even with an asymmetric prior.

Finally, further optimisation of the new pipeline is needed before we can analyse dark sirens at the redshift accessible to third-generation GW detectors.
The much larger redshift range (up to $z \sim 7$ for a BBH with an SNR of 50 in the Einstein Telsecope, \citep{ET:2025xjr})
requires care in the choice of priors, and the much higher number of events (order $10^3-10^4$ for this SNR choice, up to $10^6$ detectable events, see \cite{ET:2025xjr}) 
increases computation time and requires substantial memory. We leave this to further work, and focus on bright sirens.

\subsection{Mock LVK O5 Data}
\label{ssec:mock data}
We test our method with mock GW events, which we generate in-house using \texttt{CosmoPyro} and adapting a detection formulation similar to \citep{Finn:1992xs}. Here we describe how we generate mock data for an LVK O5 scenario.

For the underlying redshift distribution of all mock datasets we use a volumetric prior:
\begin{align}
    p(z) = \frac{dV_{\rm c}}{dz}\frac{1}{1+z}\psi(z)
\end{align}
where $\psi(z)$ is the a Madau-Dickinson star formation rate \citep{Madau:2014bja}:
\begin{align}
    \psi(z) =  \frac{(1+z)^{\gamma}}{1+[(1+z)/(1+z_{\rm p})]^{\gamma+\kappa}} \, .
\end{align}
We simulate our data using $\gamma = 2.0$, $\kappa = 3.0$, and $z_{\rm p} = 2.0$, consistent with the findings in \cite{Madau:2014bja} (and more recent LVK merger rates \cite{LIGOScientific:2026ctl}). In our analyses we fix $\kappa$ and $z_{\rm p}$ as these quantities are normally not well-inferred (see Figure 13 of \cite{LIGOScientific:2021aug})
and are not degenerate with modified gravity or population parameters. The dark sirens primary masses are drawn from a MLTP distribution,  
as illustrated in Section~\ref{ssec:spectral sirens}. 
The mass ratio is sampled from a power law\footnote{Same as in the \texttt{icarogw} pipeline \citep{Mastrogiovanni:2023zbw}, to facilitate validation tests and comparisons.} $p(q|\beta_0) \propto q^{\beta_0}$, we infer $\beta_0$ alongside the MLTP parameters. 
In order to simulate realistic asymmetric uncertainties of measurements (specifically of chirp mass $\mathcal{M}$, mass ratio $q$, and GW luminosity distance $\dgw$) we model the posteriors according to Gamma distributions, described by a shape parameter $\alpha$ and a rate parameter $\lambda$. The shape parameter determines the width and asymmetry of the distribution, while the rate parameter determines the overall scale and the position of the peak. The exact position of the peak is determined by $(\alpha-1)/\lambda$. For a true parameter $x_{\rm true}$ (obtained from the previously mentioned mass and redshift distributions) we first simulate an `observed' value $x_{\rm obs}$ by drawing the observed value from $x_{\rm obs} \sim \text{Gamma}(\alpha, x_{\rm true})$, where we have set $\lambda = x_{\rm true}$ so that the Gamma function peaks close to the true value of the parameter. We then generate posterior samples by drawing from a Gamma distribution, with shape $\alpha + 1$ and rate $x_{\rm obs}$, shifting the distribution's peak relative to the observed value. The $+1$ shift in shape cancels the $-1$ offset in the mode formula $(\alpha-1)/\lambda$, ensuring the posterior mode is centered close to $x_{\rm obs}$. We scale $\alpha$ depending on the confidence with which we constrain these parameters in parameter estimation of real data \citep{LIGOScientific:2025slb, LIGOScientific:2026wfs}, where the relative error $\approx 1/\sqrt{\alpha}$. We use average errors in chirp mass, mass ratio, and GW distance of $\sim 3\%$, $\sim 14 \%$, and $\sim 20 \%$ respectively.

We select events with an SNR $\geq 12$, utilising a computationally light 
approximated optimal SNR $\rho$ based on LVK O5 (fifth observing run) detector sensitivity:
\begin{equation}
\label{eq:SNR}
    \rho = \frac{\mathcal{A}\mathcal{M}^{5/6}\mathcal{P}I}{\dgw} \, ,
\end{equation}
where we include the binary parameters chirp mass $\mathcal{M}$, distance $\dgw$, and inclination (contained within $I$, a combination of the two polarisations of the signal) as in the standard computation of optimal SNR, see e.g. \citep{Flanagan:1997sx}. We then approximate the detector response and frequency band limitations via the bandwidth penalty $\mathcal{P}$, which suppresses the SNR when the total mass of a binary pushes the signal to frequencies below the capabilities of the detector. Physical constants (such as the gravitational constant $G$, and speed of light $c$), additional antenna response terms and other scaling factors appearing in the standard SNR computation are absorbed into the catch-all amplitude term $\mathcal{A}$. With this setup, the maximum distance at which we can observe a face-on BNS event is $\sim550$ Mpc, in line with O5 \citep{KAGRA:2013rdx}\footnote{Note that the term `range' as reported in LVK papers refers to the \textit{sky-averaged} range, to find maximum detectable distance for a network one has to multiply the reported range by 2.264, see \cite{Dominik:2014yma}}.

We test the robustness of our method for a range of four non-GR cosmologies. The distinction among the four is in the injected $\dgw/d_{\rm EM}$ function, while we fix the Hubble constant to $H_0 = 70~\rm{km \,s}^{-1}\rm{Mpc}^{-1}$ and the matter energy density to $\Omega_{\rm m} = 0.3$ for all. We generate four mock datasets of 500 events each, according to these using the same mass and redshift distributions across the four cosmologies. As we wish to test the robustness of our method against various forms of $r(z)$, we generate the modified cosmologies using both monotonic and non-monotonic distance ratios:
\begin{subequations}
\label{eq:ratios}
\begin{align}
    &r_{\rm wiggly}(z) = 1 + A \times \arcsin(kz) \,, \label{eq:ratio wiggly} \\ \nonumber&\hspace{4.5cm} A = 0.09, \,k = 4.0\,,\\
    &r_{\rm bump}(z) = 1 + A\times\exp\left(-\frac{1}{2}\left(\frac{z-z_*}{\sigma}\right)^2\right) \, , \label{eq:ratio bump}\\ \nonumber&\hspace{4.5cm}A = 0.09,\, z_* = 0.6,\, \sigma = 0.09\,,\\
    &r_{\rm damped}(z) = 1 + \exp(-\ell  z) \times A \times \sin(k z) \,, \label{eq:ratio damped}\\ \nonumber&\hspace{4.5cm} A = 0.08, \, k = 8.0, \, \ell = 1.0\,,\\
    &r_{\rm c_{\rm M}} (z) = \exp\bigg(\frac{c_{\rm M}}{2\Omega_{\Lambda 0}}\int_0^z \frac{\Omega_\Lambda(z') \dd z'}{(1+z')}\bigg) \, ,\label{eq:ratio cM}\\ \nonumber &\hspace{4.5cm}  c_{\rm M} = -0.5 \,.
\end{align}
\end{subequations}
The \textit{bump} model has a single peak at redshift $z_*$ with width $\sigma$. The \textit{wiggly} and \textit{damped} models have oscillatory features with amplitude $A$ and angular frequency $k$. The latter decays according to a damping rate $\ell$. The last ratio is a form which was mentioned in the introduction as it is a widely used phenomenological ansatz following the $\alpha_i$ formalism \citep{Bellini:2014fua}, which produces a monotonic function. The specific values of the parameters regulating these functions are chosen to be consistent with current constraints of the ratio \citep{LIGOScientific:2026uyd}.

\begin{figure*}
    \centering
        \hspace*{-0.5cm}
        \includegraphics[trim={0.3cm 0 0.2cm 0},clip,width=0.7\linewidth]{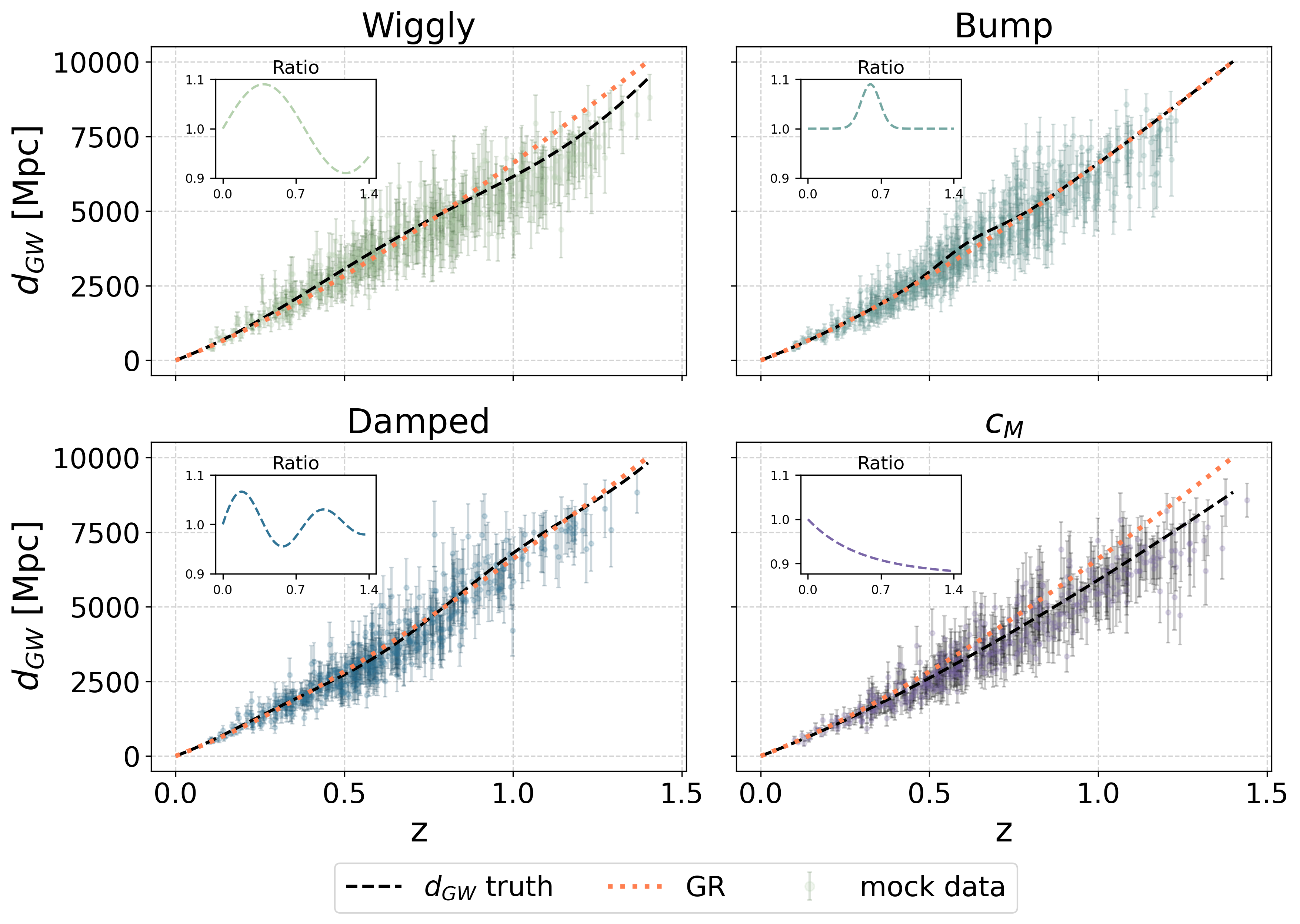}
    \caption{Visual representation of the four LVK-like mock datasets. Each panel shows GW luminosity distance as a function of redshift, each data point shows the median observed distance and the 1$\sigma$ error. We plot $\dgw$ for each fiducial non-GR cosmology in black, while the orange dotted line is the same quantity in GR. The boxes titled `ratio' within each panel show the $r(z)$ function corresponding to each cosmology (remember that GR has $r(z) = 1$) as detailed in Eqs. \eqref{eq:ratios}. 
    }
    \label{fig:data visualisation LVK}
\end{figure*}

Figure~\ref{fig:data visualisation LVK} is a visualisation of the four non-GR datasets \textit{wiggly}, \textit{bump}, \textit{damped}, and \textit{$c_M$}. Each panel displays the GW distances of the 500 mock events with corresponding distance uncertainties as a function of redshift. We show the error bars in distance since these contribute to the majority of the error budget in our ratio reconstructions The rest of the uncertainti 

The black dashed line represents the true GW distance, which deviates from the distance in GR, the orange dotted line. The inset plots show the corresponding ratios for each cosmology as a function of redshift. As mentioned above, the deviations from GR are not extreme, reaching $\sim 10\%$ as a maximum in line with current constraints \cite{LIGOScientific:2026uyd}. 

\subsection{Mock ET data}
\begin{figure*}
\centering
        \hspace*{-0.5cm}
        \includegraphics[trim={0.3cm 0 0.2cm 0},clip,width=0.7\linewidth]{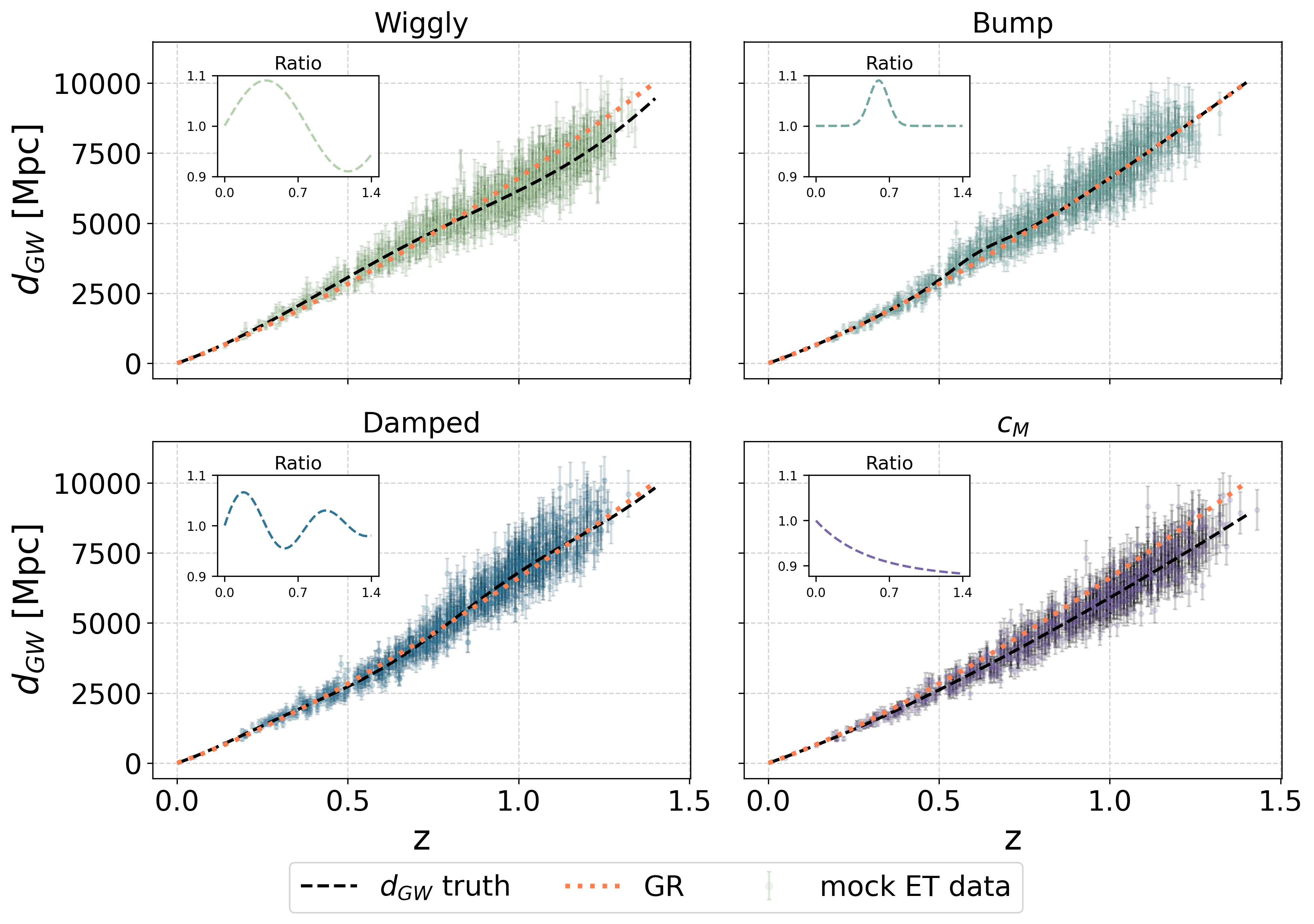}
    \caption{The 1000 mock bright siren events for each of the four modified gravity models. As with the LVK data, we show GW distance as a function of redshift, plotting the $1\sigma$ error bars in distance for each event. The black dashed lines depict the true GW distance for each case, while we also plot the corresponding GR distance as the dotted orange line. The boxes once again show the ratio $r(z)$ for each cosmology (see Eqs. \ref{eq:ratios}). 
    }
    \label{fig:data visualisation ET}
\end{figure*}

We generate 1000 events for the next generation detector case, namely the Einstein Telescope (ET) \citep{ET:2025xjr}, a ground-based detector planned for the next decade\footnote{We choose 1000 to be conservative, ET predictions claim numbers an order of magnitude larger.}. 
As previously stated, we use bright sirens for this sceneario, which will increase the accuracy of the reconstruction of $r(z)$ since we obtain a precise distance-redshift constraints due to host galaxy identification. We make this choice for simplicity (lower computational costs) and to showcase the full capabilities of the method.
We sample their masses from a Gaussian centred around $1.5 M_\odot$ and a standard deviation $\sigma=0.2$ \citep{Kiziltan:2013oja, LIGOScientific:2020aai}. 
For the distance error, the concentration in Gamma distribution is rescaled to account for the higher sensitivity of ET (10x the LVK sensitivity on average, see \cite{Gupta:2023lga}) which translates to more accurate parameter estimation, especially at higher SNRs. At leading order errors in distance and chirp mass both roughly scale as $\sim 1/\rm{SNR}$ \citep{Cutler:1994ys}. We select a threshold SNR of 12 again, computing it as presented in Eq.~\ref{eq:SNR}, but modifying the penalty window $\mathcal{P}$ to reach a frequency of 5 Hz as per ET design \citep{ET:2025xjr} (in contrast with the LVK 20 Hz limit), and adding the 10x sensitivity in the amplitude term $\mathcal{A}$. This results in average errors of $\sigma_\mathcal{M}\sim 1\%$, $\sigma_q\sim 2.8\%$, and $\sigma_{\dgw} \sim 8.3\%$. The improvement in chirp mass and mass ratio measurements is twice the one in distance, as BNS events have a smaller mass prior due to astrophysics (and are conservative compared to estimates from \cite{Maggiore:2024cwf}). We show the data points in Figure~\ref{fig:data visualisation ET}. 
We assume EM follow-up provides spectroscopic redshift measurements for all events (thanks to future observatories and EM follow-up programs, see \cite{Steeghs:2021wcr, Keinan:2024gai, Nicholl:2024ttg}), which allows us to ignore errors in $z$ as they are sub-leading to the distance errors.

\subsection{GWTC-5 Data}
\label{ssec:gwtc-5 data}
Our final dataset comprises the events from the fifth data release (GWTC-5) \citep{LIGOScientific:2026wfs} of the LVK collaboration. We use the subset of 231 binary black hole events used in the latest LVK cosmology results \citep{LIGOScientific:2026uyd}. 
The paper includes 236 events, the additional 5 have been identified as involving at least one neutron star. We do not use these as we have decided to focus on BBHs in this case in order to use the MLTP mass distribution. This aids the comparison of our validation with the mock data.
These mergers are a subset of the full GWTC-5 catalog only containing the highest significance events (FAR$< 0.25~yr^{-1}$).

\section{Results}
\label{sec:results}
\subsection{Mock LVK O5 data}
\label{ssec:mock LVK results}
We present here the results obtained with the 500
mock LVK O5 events.
\begin{figure*}
    \centering
        \includegraphics[width=0.7\linewidth]{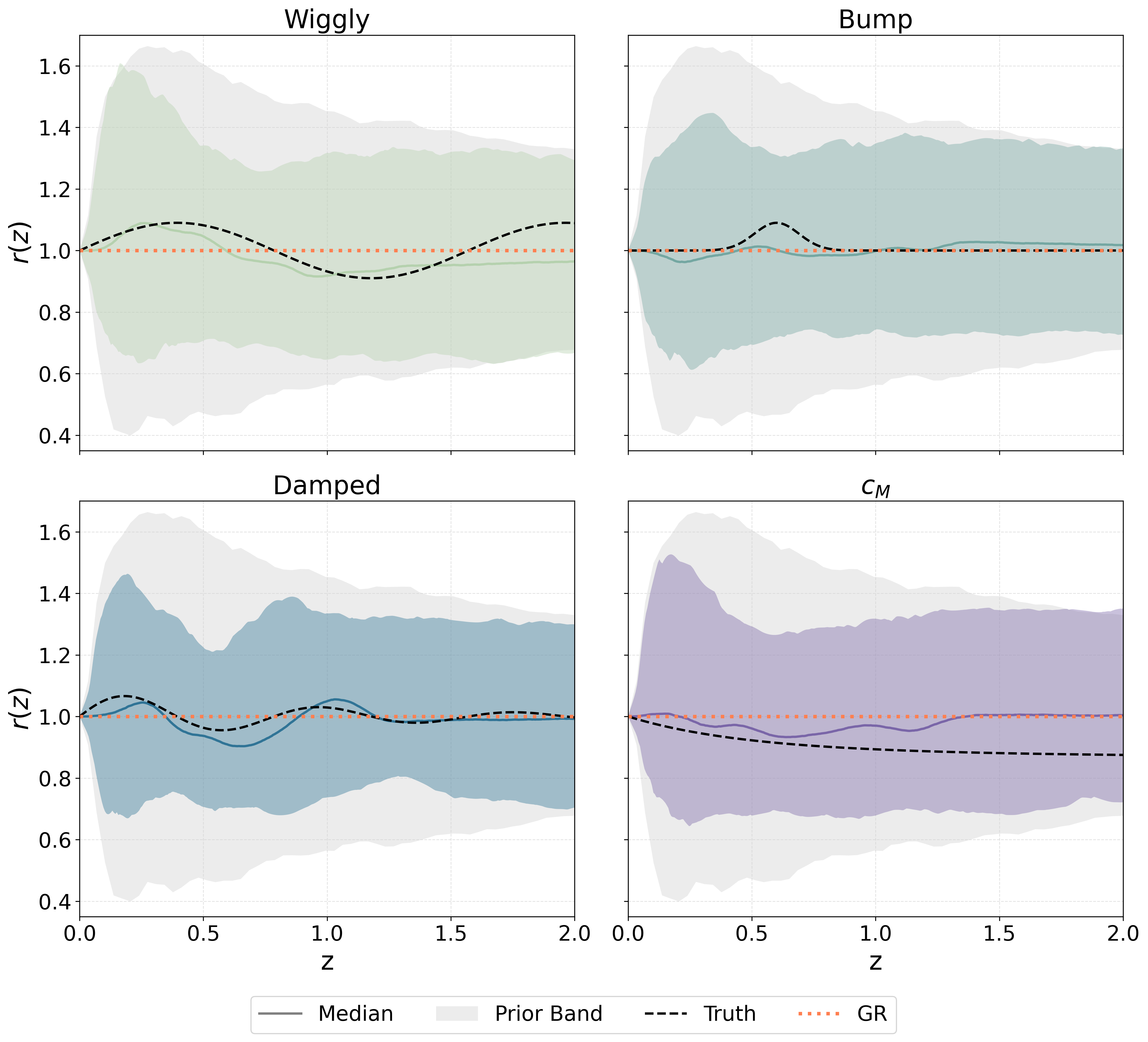}
    \caption{Reconstruction of the ratio $r(z)$ using the mock LVK O5 data as a function of redshift for the four non-GR cosmological models explored \eqref{eq:ratios}. The coloured shaded region represents 3$\sigma$ contour of our posterior, with the solid lines showing the median of each reconstruction. We also highlight the priors region on the ratio in gray. The black dashed line represents the truth for each cosmology, while the horizontal dotted line at $r=1$ corresponds to GR.}
    \label{fig:mock LVK results}
\end{figure*}
\begin{figure}
    \centering
        \hspace*{-0.5cm}
        \includegraphics[trim={0.3cm 0 0.2cm 0},clip,width=1.07\columnwidth]{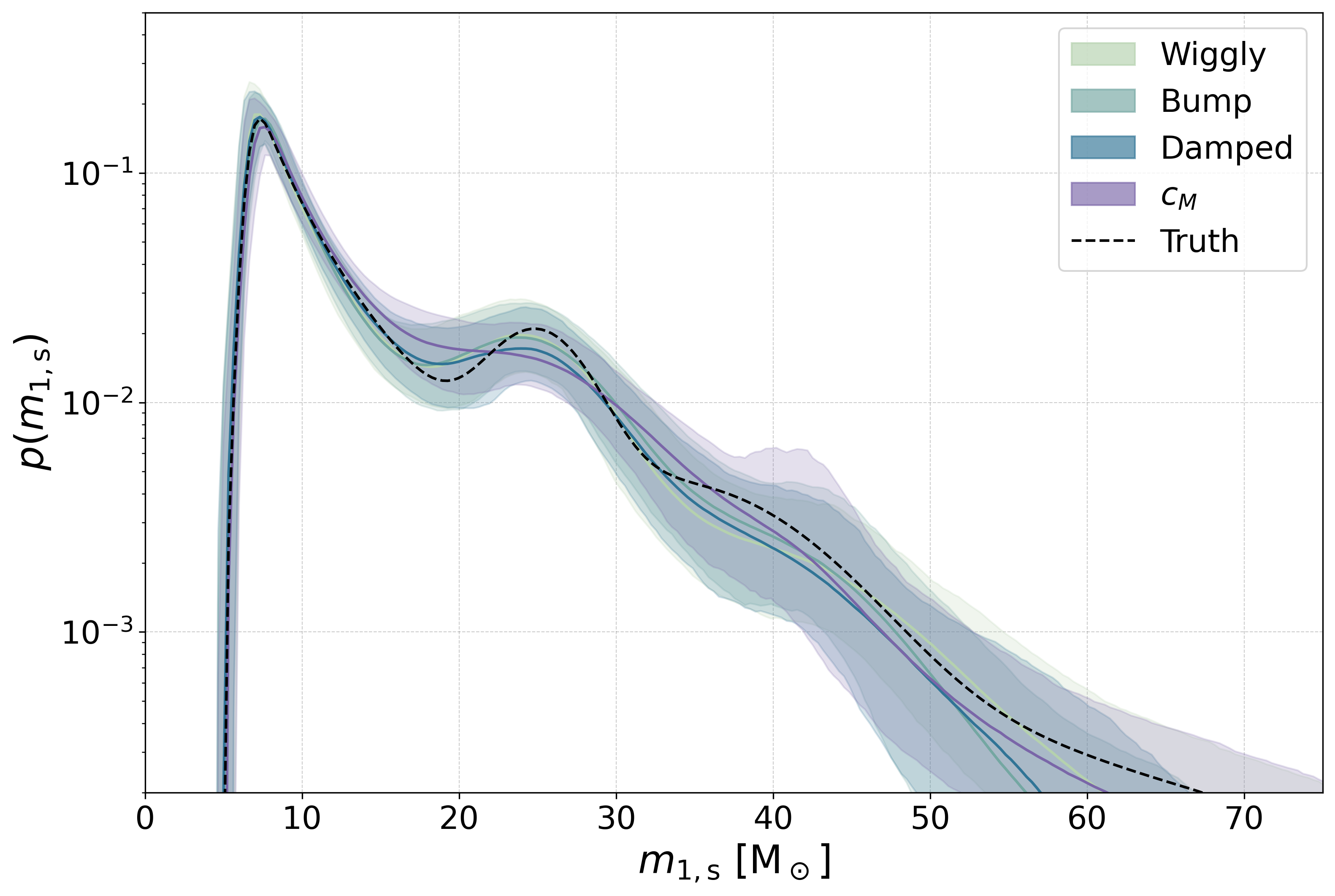}
    \caption{Reconstruction of the BBH primary mass distribution in source frame. We show the 3$\sigma$ contour of all four cosmologies in the shaded regions, obtained using the MLTP parametrisation. The black dashed line is the true distribution from which we sampled our mock data: all posteriors are consistent with it.}
    \label{fig:mass dist LVK mock}
\end{figure}
Figure~\ref{fig:mock LVK results} shows the reconstructed distance ratio $r(z)$ for the four non-GR cosmologies. In each panel we plot the reconstructed ratio's 3$\sigma$ contour (coloured), with the median plotted as the solid line of the same colour. We choose to show 3$\sigma$ contours as recording deviations from GR at 1$\sigma$ is not in itself statistically significant. The true ratio for each cosmology is shown as a dashed black line, and the shaded gray area indicates the prior band in ratio. The GR limit is highlighted as the dotted orange line at $r(z) = 1$. The reconstructions show that we cannot resolve the injected deviations from GR. This is to be expected, as we are using events with large errors in GW distance, as illustrated in Fig.~\ref{fig:data visualisation LVK}. We do observe that the data informs the posteriors, and all four reconstructions agree with the true ratios within 1$\sigma$ in the redshift range where data is present. At higher redshifts, where we have no data points, the posteriors are completely dominated by the prior. The more prominent features such as the fluctuations in $r_{\rm wiggly}$ and $r_{\rm damped}$ are imprinted in the posteriors, though not resolved well enough. The small gaussian peak from the $r_{\rm bump}$ cosmology is not strong enough to be resolved by this data. 
Unlike the other models, the $c_{\rm M}$ parametrisation is monotonic. Even though the uncertainties are too large to distinguish it from GR, the median of the reconstruction successfully tracks the downward trend before converging again at 1 at higher redshifts (as our model prescribes). Hence, despite using an oscillatory basis, no strong artificial features are introduced.

We also show the reconstruction of the source frame mass distribution in Figure~\ref{fig:mass dist LVK mock}. We show the 3$\sigma$ contour result from all four cosmologies in the shaded coloured regions.
We see that the distribution is recovered well, with the truth within the 3$\sigma$ contour for all cosmologies. This highlights the strength of our method: introducing a modified gravity analysis does not impact our abilities to infer the mass distribution. We report all inferred values for each parameter of the MLTP distribution in Appendix~\ref{appendix:LVK extra}. We highlight here the inferred values for the distribution peaks, found in the correct position for all cosmologies 
. 
The median of the higher peak is found around $\mu_{g,\mathrm{high}} \sim 36.7-38.6M_\odot$ (true $\mu_{g,\mathrm{high}} = 35M_\odot$, with average error of $13\%$ ($68.3\%$ confidence interval). The median of the lower peak is consistently localised between $\mu_{g,\mathrm{low}} \sim 23.7-24.4 M_\odot$ (true $\mu_{g,\mathrm{low}} = 25M_\odot$) with an average 1$\sigma$ error of $5\%$. For reference, the latest LVK results find a peak around the same region $\sim 27 M_\odot$ using the MP model, and report a $10\%$ error.

On a final remark, we want to highlight the average 3 (1)$\sigma$ error on the ratio for all four cosmologies is 24 (10)\% for the plotted redshift range (up to $z=2$).
To reach errors a factor of 10 smaller at this detector sensitivity and $\dgw$ posterior precision we require a factor of 100 times more events (so 50000), since our errors roughly scale as $1/\sqrt{N}$, for $N$ data points. While this number is most likely not attainable with the current LVK network \citep{LIGOScientific:2026ctl}, ET is predicted to see $10^5-10^6$ events per year. The issue then will be improving the computational efficiency of the pipeline to undertake such a large task. 

\subsection{Mock ET data}
\label{ssec:ET results}
In this section we present results for the ET detector mock bright sirens. 
\begin{figure*}
    \centering
    \includegraphics[width=0.7\linewidth]{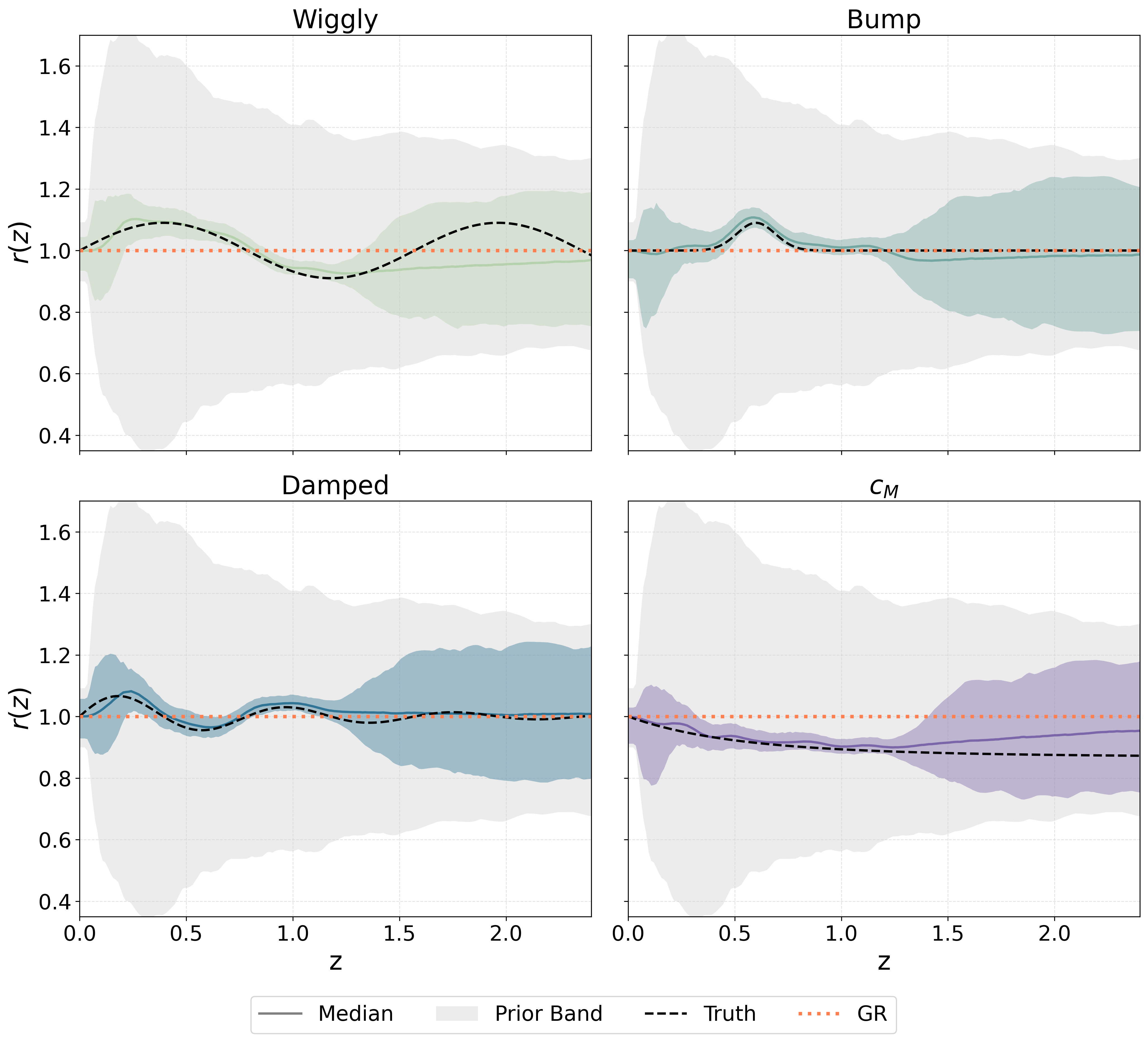}
    \caption{Reconstruction of the ratio $r(z)$ as a function of redshift using ET mock data. The plot is analogous to Figure~\ref{fig:mock LVK results}. Once again, the coloured shaded region represent 3$\sigma$ confidence interval of the posteriors, while we show the priors in gray. Here as well the solid lines correspond to the median of the reconstruction, the dashed correspond to the underlying truth for each modified gravity model, and GR is the orange dotted line. All the contours disagree with GR, to different extents which depend on shape and sharpness of the features in the true ratio. 
    }
    \label{fig:ET ratio}
\end{figure*}
Figure~\ref{fig:ET ratio}, analogous to the one in the previous section, shows the reconstruction of $r(z)$ for the four cosmologies. Once again we plot the 3$\sigma$ confidence interval of the reconstruction in the shaded coloured regions, while the gray shows the prior band. The GR line is again highlighted in orange, with the truth being the dashed black line. These plots show the true capabilities of our reconstruction: not only are we able to recover the true shape of the underlying ratio in each cosmology, we also observe the 3$\sigma$ contour is now showing the inconsistency with GR for all four cases. Hence, we are able to reconstruct both oscillating, and monotonic ratios, including ratios with sudden smaller features such as the $r_{\rm bump}$ cosmology on the top right. The posterior reduces back to the prior where no data is present. This shows that with enough high quality data points we can distinguish between GR and non-GR using the ratio of GW to EM distance with our weakly-modelled approach, without needing to resort to parametrised ansatzes. 
The average 3 (1)$\sigma$ error on the ratios is 9 (4)\%. The number of events is only twice that of the LVK mock catalog, but the results on $r(z)$ in are $\sim 2.5$ times better due to our usage of bright sirens due to the simplified form of the likelihood in Eq.~\ref{eq:bright likelihood}.
We have seen that sirens provide precise reconstructions of the ratio in the low-redshift range. However, additional information remains to be studied at higher redshifts with dark sirens from next generation detectors. We conclude that the bright and dark siren methods should be used jointly to constrain $r(z)$ over a wide redshift range. 

\subsection{GWTC-5 Data} 
\label{ssec:real data results}
\begin{figure*}
    \makebox[\textwidth][c]{
        \begin{minipage}{1.1\textwidth}
            \centering
            \begin{subfigure}[b]{0.5\linewidth}
                \centering                \includegraphics[width=\linewidth, trim={0cm 0cm 0cm 1cm}]{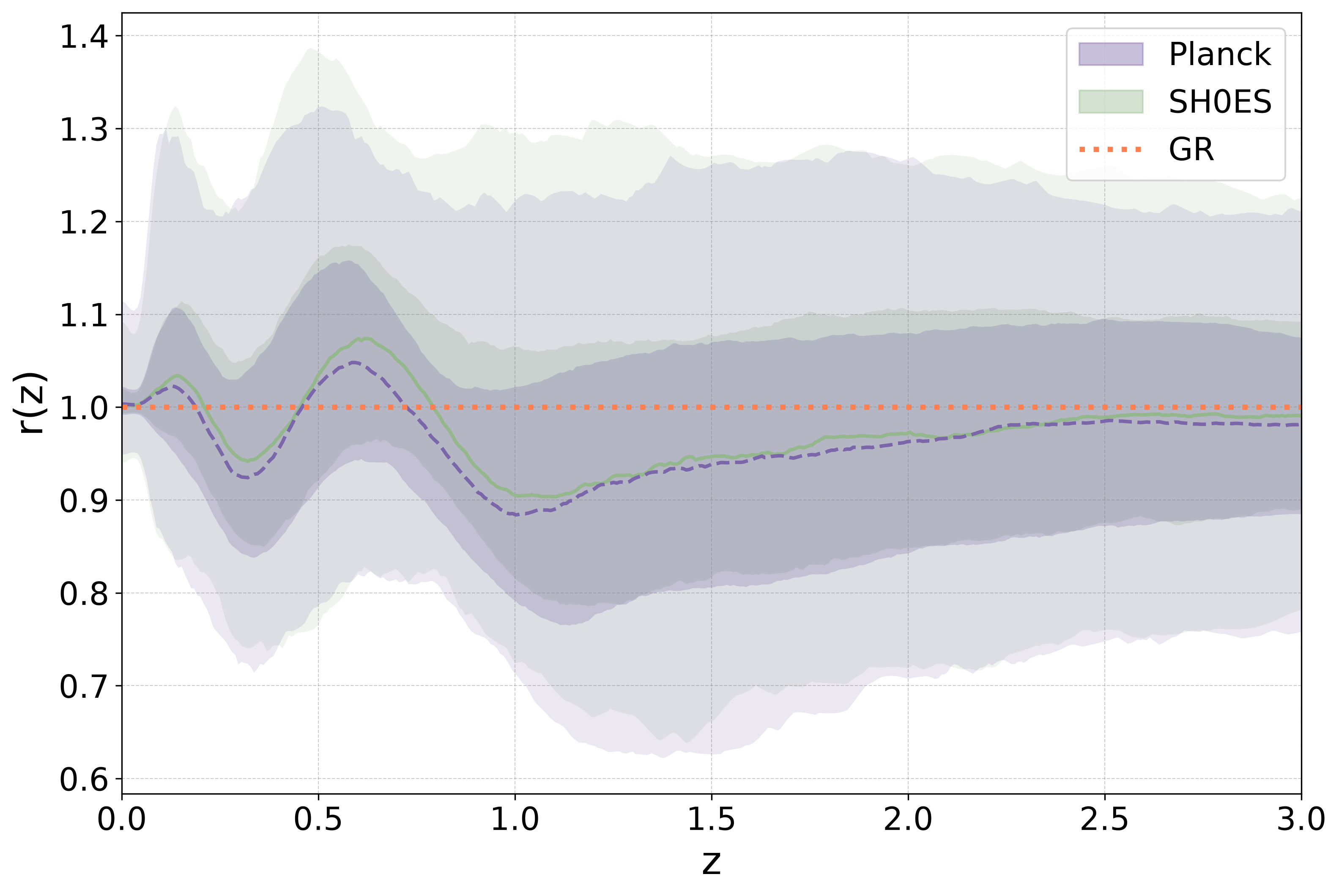}
                \caption{}
                \label{fig:GWTC-5 ratio}
            \end{subfigure}%
            \begin{subfigure}[b]{0.5\linewidth}
                \centering
                \includegraphics[width=\linewidth, trim={0cm 0.1cm 0cm 0cm}]{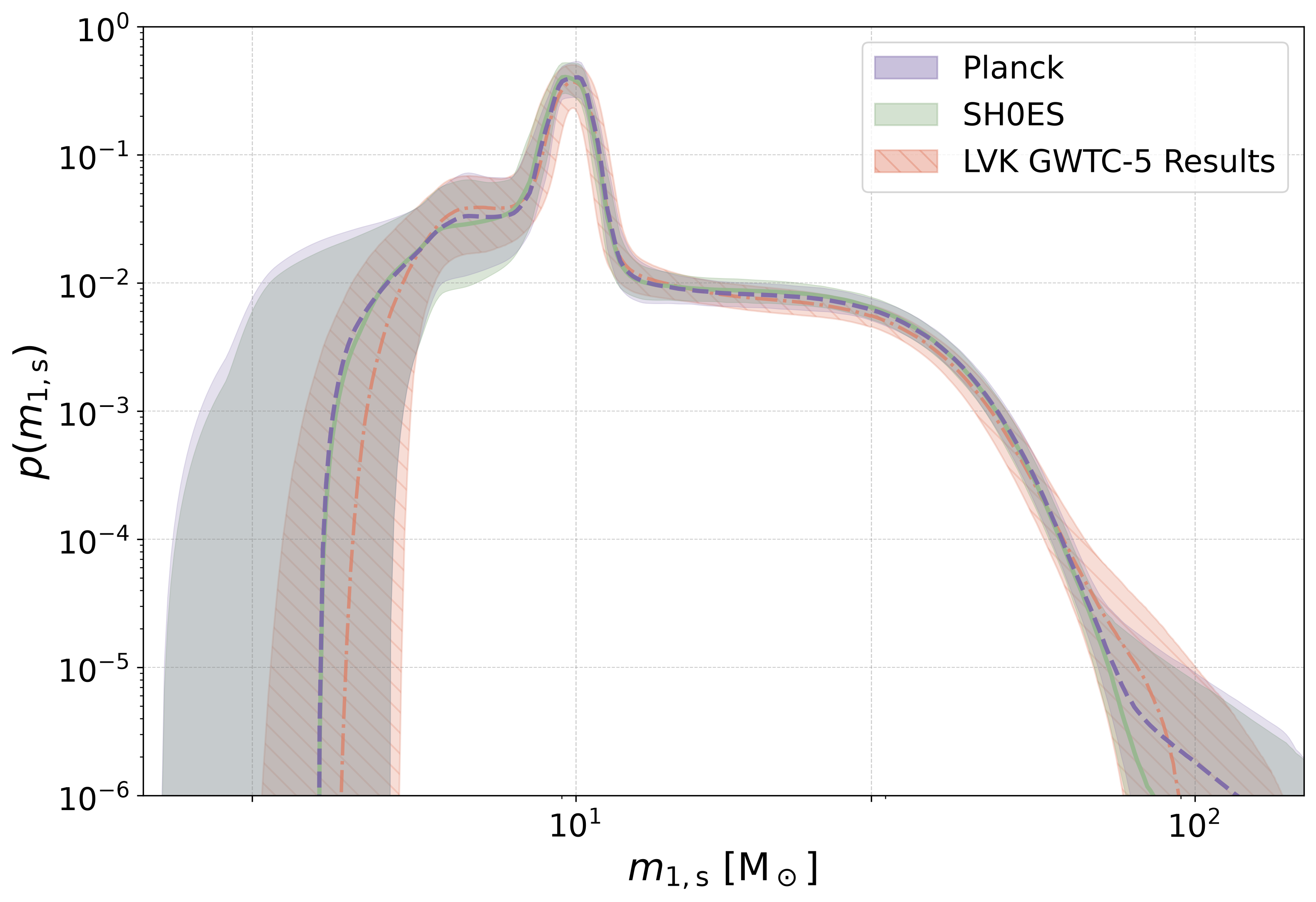}
                \caption{}
                \label{fig:GWTC-5 mass dist}
            \end{subfigure}
        \end{minipage}
    }
    \caption{Results for GWTC-5, the latest LVK data release. We plot the reconstructed ratio as a function of redshift in left panel. We show 1 and 3$\sigma$ contours for both Planck (purple, dashed line median) and SH0ES (green, solid line median) values of $H_0$. The orange dotted line is GR. The panel on the right shows the posterior on the primary mass distribution for both cosmologies, with the latest LVK results overlayed in orange.}
    \label{fig:GWTC-5 results}
\end{figure*}

Finally, we present results using real data of 231 BBHs from GWTC-5. We once again fix the background cosmology to isolate the modified gravity signatures from degeneracies. Given the current disagreement on the cosmic expansion rate, we fix $H_0$ to two values to perform our analyses: the first is the cosmic microwave background value given by Planck of $67.4~\rm{km s}^{-1}\rm{Mpc}^{-1}$ \citep{Planck:2018vyg}, and the second is the local measurement by the SH0ES collaboration, $73.04~\rm{km s}^{-1}\rm{Mpc}^{-1}$ \citep{Riess:2021jrx}. The resulting reconstructions are shown in Figure~\ref{fig:GWTC-5 results}. In Figure~\ref{fig:GWTC-5 ratio} we plot results for both Planck and SH0ES $H_0$ values, in purple and green respectively. We show 1 and 3$\sigma$ contours in the respective colours and the medians of the two reconstructions are the dashed (Planck) and solid (SH0ES) lines. The orange dotted line denotes the GR expectation of $r(z)=1$. 
Although both reconstructions exhibit minor oscillatory structure, both cosmological backgrounds yield results that are fully consistent with GR. Additionally, Figure~\ref{fig:GWTC-5 mass dist} displays the 68\% confidence interval of the primary mass distribution $p(m_{1, \rm s})$ for both Planck and SH0ES scenarios. For comparison, we plot the latest LVK results for the MLTP distribution in the shaded orange region (from \cite{LIGOScientific:2026uyd}). Our inferred mass distributions are in excellent agreement with the official LVK findings, further confirming that our weakly-modelled modified gravity framework does not produce any unphysical shifts in the population distribution. Additionally, while LVK analyses vary $H_0$, we only vary the MG parameters (fixing $H_0$) and find comparable error bars in $p(m_{1, \rm s})$ which further demonstrates the statistical robustness of our method.
The only mild discrepancy occurs at low mass, due to the difference in how smoothing is applied around $m_{\rm min}$ in our pipeline compared to the LVK pipelines: whereas \texttt{icarogw} produces a sharp drop to zero, \texttt{CosmoPyro} applies a gentle smoothing with a broader peak of the power law, allowing for a slower tapering off (hence the broader band).

\section{Conclusions}
\label{sec:conclusions}
We have presented a new, weakly-modelled method to detect deviations from GR by reconstructing the ratio $r(z) = \dgw/d_{\rm EM}$ between the distance measured from gravitational wave signals and the standard EM luminosity distance. By decomposing the ratio into a Fourier series we are able to reconstruct a wide family of functional forms of $r(z)$, both monotonic and non-monotonic, and with either periodic or irregular features. Our method avoids the restricted theoretical assumptions that limit current parametric analyses. In doing so, we widen our ability to capture or eliminate diverse signals of deviation from GR on cosmological scales.

In Section~\ref{ssec:mock LVK results} we used the spectral sirens method with 500 LVK O5 mock events and a fixed cosmology to obtain a broad posterior for $r(z)$ with 1$\sigma$ average errors of 10\%. We however show that the results are unbiased and expected for typical O5 events. We also demonstrated the full capabilities of our reconstructions in Section~\ref{ssec:ET results}, where a next-generation mock data set (1000 bright sirens) enabled recovery of the injected deviations from GR to 3$\sigma$ significance. Finally, to demonstrate our method is ready to be deployed with real data, we applied our method to 231 binary black hole events from the GWTC-5 catalog in Section~\ref{ssec:real data results}. We do not observe any deviations from GR, and obtain a posterior on the source frame mass distribution that is fully consistent with that of \cite{LIGOScientific:2026ctl} and \cite{LIGOScientific:2026uyd}. This again demonstrates our method does not bias or introduce strong degeneracies with population inference.

Looking ahead, we wish to highlight a few avenues for future work. 
While our reconstruction is highly flexible, we have not explored here how varying the power spectrum parameters affects our results. Though we have made physically motivated choices in order to fix their values, future work could treat these as free parameters as well. However, this would require a larger number of precisely measured events in order to obtain constraints on par with what we have shown.

To mitigate potential mismodelling of the mass distribution -- a potential risk for all spectral/dark siren analyses -- we wish to expand this analysis to non-parametric forms of the mass distribution. 
This feature is implemented in \texttt{CosmoPyro}\footnote{\url{https://github.com/KonstantinLeyde/CosmoPyro}}: as described in detail in \cite{COSMOPYRO}, the pipeline allows for a joint gaussian process reconstruction of the primary mass and the mass ratio distributions. Performing a full weakly modelled analysis of population and modified gravity would provide a (though weaker) more robust result.

Though bright sirens have shown to provide a precise result at low redshifts ($\lesssim 1.5$), we must wait until next generation detectors to see this type of constraint. To increase precision with current data, multiple methods for constraining the ratio $r(z)$ could be combined. Namely, this would entail adding galaxy catalog information, either with a line-of-sight redshift prior or with cross-correlations of GW and galaxy maps. 
When carrying out a joint analysis any covariance between the methods must be carefully assessed. This may be non-trivial for the spectral sirens and line-of-sight redshift prior; it is likely to be low for spectral sirens and GW-galaxy cross-correlation, as mass distribution information is not directly present in the latter. 

Additionally, different EM tracers could enter the analysis alongside GWs. Examples are extending current $H_0$ inference methods with radio sources (see e.g. \cite{Zazzera:2025ord} and \cite{Dupletsa:2026uqs}) to test GR, or cross-correlating GWs with supernovae \textcolor{blue}{(Colangeli E., Santiago de Matos I., Baker T., in prep.)} for a background-independent test of the $d_{\rm EM} - \dgw$ relation. 

Furthermore, one should also combine these GW-EM results with EM-only constraints: alternative theories of gravity which modify the GW signal often also affect the growth rate of structure, weak lensing, and clustering (e.g. this occurs in Horndeski theory \cite{Horndeski:1974wa}). A full, consistent analysis would require non-parametric inference of these EM observables as well. Some foundational work has been made by \cite{Pogosian:2021mcs} and \cite{Raveri:2021dbu} to utilise large-scale structure to constrain deviations from GR in a non-parametric way.

Finally, throughout this work we have assumed the background expansion history is given by the $\Lambda$CDM model, and have not addressed the new evidence pointing towards a dynamical dark energy scenario from \cite{DESI:2025zgx}. If this preference persists, analyses of MG effects in GW propagation should be extended to account for the dark energy equation of state as well. 

\section*{Acknowledgements}
We thank Antonio Enea Romano, Johannes Noller, and Wolfgang Enzi for useful discussions. We additionally thank Matteo Tagliazucchi for  helpful comments and carrying out LVK internal review. E.C. and T.B. are supported by ERC Starting Grant SHADE (grant no.\,StG 949572). T. B. is further supported by a Royal Society University Research Fellowship (grant no.\,URF$\backslash$R$\backslash$231006). A.C. is supported by the China Postdoctoral Science Foundation under Grant No. 2025M773325, and the National Natural Science Foundation of China (NSFC) under Grant No. E414660101 and 12147103. 
Part of the computations were performed on the \texttt{Sciama} High Performance Computing (HPC) cluster, which is supported by the Institute of Cosmology and Gravitation (ICG), the South-East Physics Network (SEPNet) and the University of Portsmouth.
The computations reported in this paper were (in part) performed using resources made available by the Flatiron Institute.
The Center for Computational Astrophysics at the Flatiron Institute is supported by the Simons Foundation. This material is based upon work supported by NSF's LIGO Laboratory which is a major facility fully funded by the National Science Foundation.
\bibliographystyle{mnras}
\bibliography{main} 

\appendix

\section{LVK mock data: full results}
\label{appendix:LVK extra}
We provide here additional results from our runs. In Table~\ref{tab:mock LVK results} we report the posteriors on all black hole mass distribution parameters and the redshift parameter $\gamma$ for the four non-GR datasets used in this work. We also list the injected value for each parameter. 
\begin{table}
\centering
\caption{Mock LVK results for the different non-GR cosmologies.}
\label{tab:mock LVK results}
\renewcommand{\arraystretch}{1.2}
\setlength{\tabcolsep}{3.5pt}
\footnotesize

\begin{tabular}{lccccc}
\hline\hline
\textbf{Parameter} & \textbf{Truth} & \textbf{wiggly} & \textbf{bump} & \textbf{damped} & \textbf{$c_{\rm M}$} \\
\hline
$\alpha$ & $3.1$ & $3.42^{+0.46}_{-0.31}$ & $3.85^{+0.46}_{-0.37}$ & $3.45^{+0.52}_{-0.45}$ & $3.34^{+0.29}_{-0.24}$ \\
$\beta_0$ & $1.2$ & $1.35^{+0.21}_{-0.21}$ & $1.30^{+0.21}_{-0.21}$ & $1.44^{+0.21}_{-0.22}$ & $1.51^{+0.22}_{-0.22}$ \\
$\delta_m$ & $3.0$ & $2.84^{+0.99}_{-0.84}$ & $3.85^{+0.71}_{-0.86}$ & $3.31^{+1.00}_{-0.96}$ & $3.53^{+0.91}_{-1.06}$ \\
$\lambda_g$ & $0.1$ & $0.11^{+0.03}_{-0.02}$ & $0.10^{+0.03}_{-0.02}$ & $0.09^{+0.03}_{-0.02}$ & $0.11^{+0.05}_{-0.03}$ \\
$\lambda_{g,\mathrm{low}}$ & $0.7$ & $0.83^{+0.06}_{-0.10}$ & $0.82^{+0.08}_{-0.14}$ & $0.84^{+0.09}_{-0.15}$ & $0.88^{+0.08}_{-0.16}$ \\
$m_{\max}$ & $70\,M_\odot$ & $94.60^{+64.00}_{-19.01}$ & $104.03^{+64.75}_{-31.75}$ & $66.47^{+71.19}_{-7.91}$ & $92.03^{+62.57}_{-15.60}$ \\
$m_{\min}$ & $5\,M_\odot$ & $4.99^{+0.22}_{-0.23}$ & $4.95^{+0.20}_{-0.20}$ & $4.90^{+0.22}_{-0.21}$ & $5.05^{+0.22}_{-0.21}$ \\
$\mu_{g,\mathrm{high}}$ & $35\,M_\odot$ & $38.58^{+4.39}_{-4.76}$ & $37.11^{+4.99}_{-4.70}$ & $36.72^{+5.21}_{-4.71}$ & $38.24^{+4.95}_{-6.03}$ \\
$\mu_{g,\mathrm{low}}$ & $25\,M_\odot$ & $24.41^{+1.09}_{-1.09}$ & $24.19^{+1.13}_{-1.23}$ & $24.40^{+1.12}_{-1.39}$ & $23.73^{+1.84}_{-2.54}$ \\
$\sigma_{g,\mathrm{high}}$ & $7.0$ & $8.22^{+1.28}_{-2.66}$ & $7.45^{+1.68}_{-2.85}$ & $6.72^{+2.14}_{-2.92}$ & $6.59^{+2.32}_{-3.83}$ \\
$\sigma_{g,\mathrm{low}}$ & $3.0$ & $4.03^{+0.88}_{-0.77}$ & $4.59^{+1.20}_{-1.07}$ & $4.23^{+1.77}_{-1.06}$ & $6.20^{+2.46}_{-1.97}$ \\
$\gamma$ & $2.0$ & $2.75^{+0.73}_{-0.76}$ & $2.77^{+0.73}_{-0.73}$ & $2.91^{+0.68}_{-0.74}$ & $2.95^{+0.67}_{-0.76}$ \\
\hline\hline
\end{tabular}
\end{table}
\section{Mass population priors}
We provide the priors used for both mock and GWTC-5 dark sirens in pur analyses. Any discrepancy in values is due to the difference in underlying mass distribution between the mock data and the real data. We purposefully simulated the mock data using more exaggerated features to show the robustness of our method.
\begin{table}
\centering
\caption{Priors across both analysis configurations.}
\label{tab:priors_single_column}
\renewcommand{\arraystretch}{1.15}
\setlength{\tabcolsep}{5pt}
\small
\begin{tabular}{lcc}
\hline\hline
\textbf{Parameter} & \textbf{Mock Data} & \textbf{GWTC-5} \\
\hline
\multicolumn{3}{l}{\textit{Mass Distribution}} \\
$\alpha$ & $\mathcal{U}[1.5, \, 5.0]$ & $\mathcal{U}[1.5, \, 12.0]$ \\
$\beta_0$ & $\mathcal{U}[-4.0, \, 12.0]$ & $\mathcal{U}[-4.0, \, 12.0]$ \\
$\delta_m$ & $\mathcal{U}[1.0, \, 5.0]$ & $\mathcal{U}[0.001, \, 10.0]$ \\
$m_{\min}$ & $\mathcal{U}[2.0, \, 10.0]$ & $\mathcal{U}[2.0, \, 10.0]$ \\
$m_{\max}$ & $\mathcal{U}[40.0, \, 200.0]$ & $\mathcal{U}[140.0, \, 200.0]$ \\
$\mu_{g,\mathrm{low}}$ & $\mathcal{U}[10.0, \, 30.0]$ & $\mathcal{U}[5.0, \, 15.0]$ \\
$\mu_{g,\mathrm{high}}$ & $\mathcal{U}[30.0, \, 60.0]$ & $\mathcal{U}[15.0, \, 100.0]$ \\
$\sigma_{g,\mathrm{low}}$ & $\mathcal{U}[1.0, \, 10.0]$ & $\mathcal{U}[0.4, \, 5.0]$ \\
$\sigma_{g,\mathrm{high}}$ & $\mathcal{U}[1.0, \, 10.0]$ & $\mathcal{U}[0.4, \, 15.0]$ \\
$\lambda_g$ & $\mathcal{U}[0.0, \, 1.0]$ & $\mathcal{U}[0.0, \, 1.0]$ \\
$\lambda_{g,\mathrm{low}}$ & $\mathcal{U}[0.0, \, 1.0]$ & $\mathcal{U}[0.0, \, 1.0]$ \\
\hline
\multicolumn{3}{l}{\textit{Redshift Evolution}} \\
$\gamma$ & $\mathcal{U}[-1.0, \, 4.0]$ & $\mathcal{U}[0.0, \, 5.0]$ \\
$\kappa$ & $\delta(3.0)$ & $\mathcal{U}[0.0, \, 6.0]$ \\
$z_p$ & $\delta(2.0)$ & $\mathcal{U}[0.0, \, 4.0]$ \\
\hline
\hline\hline
\end{tabular}
\end{table}

\label{lastpage}
\end{document}

%% file: macros.tex
\def \dd {\mathrm{d}} 

\newcommand{\beq}{\begin{equation}}
\newcommand{\eeq}{\end{equation}}
\newcommand{\bea}{\begin{eqnarray}}
\newcommand{\eea}{\end{eqnarray}}


\renewcommand\S{\mathcal{S}}

\newcommand\ees{\end{eqnarray}}
\newcommand\bees{\begin{eqnarray}}

\newcommand{\dgw}{d_{\rm GW}}

\newcommand{\Lpop}{\Lambda_{\rm pop}}

\newcommand{\Lm}{\Lambda_{\rm MG}}

%% file: main.bib
@article{phan2019composable,
  author  = {Phan, Du and Pradhan, Neeraj and Jankowiak, Martin},
  title   = "{Composable Effects for Flexible and Accelerated Probabilistic Programming in NumPyro}",
  journal = {arXiv e-prints},
  year    = 2019,
  eprint  = {1912.11554},
  archivePrefix = {arXiv},
  primaryClass = {stat.ML},
  adsurl  = {https://ui.adsabs.harvard.edu/abs/2019arXiv191211554P}
}

@article{bingham2019pyro,
  author    = {Eli Bingham and
               Jonathan P. Chen and
               Martin Jankowiak and
               Fritz Obermeyer and
               Neeraj Pradhan and
               Theofanis Karaletsos and
               Rohit Singh and
               Paul A. Szerlip and
               Paul Horsfall and
               Noah D. Goodman},
  title     = {Pyro: Deep Universal Probabilistic Programming},
  journal   = {J. Mach. Learn. Res.},
  volume    = {20},
  pages     = {28:1--28:6},
  year      = {2019},
  adsurl       = {http://jmlr.org/papers/v20/18-403.html}
}

@article{Planck:2018vyg,
    author = "Aghanim, N. and others",
    collaboration = "Planck",
    title = "{Planck 2018 results. VI. Cosmological parameters}",
    eprint = "1807.06209",
    archivePrefix = "arXiv",
    primaryClass = "astro-ph.CO",
    doi = "10.1051/0004-6361/201833910",
    journal = "Astron. Astrophys.",
    volume = "641",
    pages = "A6",
    year = "2020",
    note = "[Erratum: Astron.Astrophys. 652, C4 (2021)]"
}

@article{Gray:2023wgj,
    author = "Gray, Rachel and others",
    title = "{Joint cosmological and gravitational-wave population inference using dark sirens and galaxy catalogues}",
    eprint = "2308.02281",
    archivePrefix = "arXiv",
    primaryClass = "astro-ph.CO",
    doi = "10.1088/1475-7516/2023/12/023",
    journal = "JCAP",
    volume = "12",
    pages = "023",
    year = "2023"
}

@article{LIGOScientific:2021aug,
    author = "Abbott, R. and others",
    collaboration = "LIGO Scientific, Virgo, KAGRA",
    title = "{Constraints on the Cosmic Expansion History from GWTC\textendash{}3}",
    eprint = "2111.03604",
    archivePrefix = "arXiv",
    primaryClass = "astro-ph.CO",
    reportNumber = "LIGO-P2100185-v6, LIGO-P2100185-v5",
    doi = "10.3847/1538-4357/ac74bb",
    journal = "Astrophys. J.",
    volume = "949",
    number = "2",
    pages = "76",
    year = "2023"
}

@article{LIGOScientific:2025jau,
    author = "Abac, A. G. and others",
    collaboration = "LIGO Scientific, VIRGO, KAGRA",
title = "{GWTC-4.0: Constraints on the Cosmic Expansion Rate and Modified Gravitational-wave Propagation}",
      journal = {arXiv e-prints},
          year = 2025,
         month = sep,
           eid = {arXiv:2509.04348},
         pages = {arXiv:2509.04348},
archivePrefix = {arXiv},
        eprint = {2509.04348},
 primaryClass = {astro-ph.CO},
        adsurl = {https://ui.adsabs.harvard.edu/abs/2025arXiv250904348T}
}

@article{LIGOScientific:2025pvj,
    author = "Abac, A. G. and others",
    collaboration = "LIGO Scientific, VIRGO, KAGRA",
 title = "{GWTC-4.0: Population Properties of Merging Compact Binaries}",
      journal = {\apjl},
          year = 2026,
         month = jul,
        volume = {1005},
        number = {2},
           eid = {L51},
         pages = {L51},
           doi = {10.3847/2041-8213/ae771e},
archivePrefix = {arXiv},
        eprint = {2508.18083},
 primaryClass = {astro-ph.HE},
        adsurl = {https://ui.adsabs.harvard.edu/abs/2026ApJ..1005L..51A}
}

@article{LIGOScientific:2026qni,
    author = "Abac, A. G. and others",
    collaboration = "LIGO Scientific, VIRGO, KAGRA",
title = "{GWTC-4.0: Tests of General Relativity. I. Overview and General Tests}",
      journal ="",
          year = 2026,
         month = mar,
           eid = {arXiv:2603.19019},
         pages = {arXiv:2603.19019},
archivePrefix = {arXiv},
        eprint = {2603.19019},
 primaryClass = {gr-qc},
        adsurl = {https://ui.adsabs.harvard.edu/abs/2026arXiv260319019T}
}

@article{LIGOScientific:2017zic,
    author = "Abbott, B. P. and others",
    collaboration = "LIGO Scientific, Virgo, Fermi-GBM, INTEGRAL",
    title = "{Gravitational Waves and Gamma-rays from a Binary Neutron Star Merger: GW170817 and GRB 170817A}",
    eprint = "1710.05834",
    archivePrefix = "arXiv",
    primaryClass = "astro-ph.HE",
    reportNumber = "LIGO-P1700308",
    doi = "10.3847/2041-8213/aa920c",
    journal = "Astrophys. J. Lett.",
    volume = "848",
    number = "2",
    pages = "L13",
    year = "2017"
}

@article{Mastrogiovanni:2023zbw,
    author = "Mastrogiovanni, Simone and Pierra, Gr\'egoire and Perri\`es, St\'ephane and Laghi, Danny and Caneva Santoro, Giada and Ghosh, Archisman and Gray, Rachel and Karathanasis, Christos and Leyde, Konstantin",
    title = "{ICAROGW: A python package for inference of astrophysical population properties of noisy, heterogeneous, and incomplete observations}",
    eprint = "2305.17973",
    archivePrefix = "arXiv",
    primaryClass = "astro-ph.CO",
    doi = "10.1051/0004-6361/202347007",
    journal = "Astron. Astrophys.",
    volume = "682",
    pages = "A167",
    year = "2024"
}

@article{Madau:2014bja,
    author = "Madau, Piero and Dickinson, Mark",
    title = "{Cosmic Star Formation History}",
    eprint = "1403.0007",
    archivePrefix = "arXiv",
    primaryClass = "astro-ph.CO",
    doi = "10.1146/annurev-astro-081811-125615",
    journal = "Ann. Rev. Astron. Astrophys.",
    volume = "52",
    pages = "415--486",
    year = "2014"
}

@article{Belgacem:2019zzu,
    author = "Belgacem, Enis and Foffa, Stefano and Maggiore, Michele and Yang, Tao",
    title = "{Gaussian processes reconstruction of modified gravitational wave propagation}",
    eprint = "1911.11497",
    archivePrefix = "arXiv",
    primaryClass = "astro-ph.CO",
    doi = "10.1103/PhysRevD.101.063505",
    journal = "Phys. Rev. D",
    volume = "101",
    number = "6",
    pages = "063505",
    year = "2020"
}

@article{Afroz:2023ndy,
    author = "Afroz, Samsuzzaman and Mukherjee, Suvodip",
    title = "{A model-independent precision test of general relativity using bright standard sirens from ongoing and upcoming detectors}",
    eprint = "2312.16292",
    archivePrefix = "arXiv",
    primaryClass = "astro-ph.CO",
    doi = "10.1093/mnras/stae951",
    journal = "Mon. Not. Roy. Astron. Soc.",
    volume = "530",
    number = "4",
    pages = "3812--3826",
    year = "2024"
}

@article{Afroz:2024oui,
    author = "Afroz, Samsuzzaman and Mukherjee, Suvodip",
    title = "{A model-independent precision test of General Relativity using LISA bright standard sirens}",
    eprint = "2406.08791",
    archivePrefix = "arXiv",
    primaryClass = "astro-ph.CO",
    doi = "10.1088/1475-7516/2024/10/100",
    journal = "JCAP",
    volume = "10",
    pages = "100",
    year = "2024"
}

@article{Vallejo-Pena:2026tmy,
    author = "Vallejo-Pe{\~n}a, Sergio Andr{\'e}s and Romano, Antonio Enea and Gair, Jonathan",
    title = "{Non parametric constraints of gravitational-electromagnetic luminosity distance ratio}",
    eprint = "2603.25305",
    pages = "2603.25305",
    archivePrefix = "arXiv",
    primaryClass = "gr-qc",
    adsurl="https://arxiv.org/abs/2603.25305",
    month = mar,
    year = "2026", 
    journal = {arXiv e-prints}
}

@article{Mandel:2018mve,
    author = "Mandel, Ilya and Farr, Will M. and Gair, Jonathan R.",
    title = "{Extracting distribution parameters from multiple uncertain observations with selection biases}",
    eprint = "1809.02063",
    archivePrefix = "arXiv",
    primaryClass = "physics.data-an",
    doi = "10.1093/mnras/stz896",
    journal = "Mon. Not. Roy. Astron. Soc.",
    volume = "486",
    number = "1",
    pages = "1086--1093",
    year = "2019"
}

@article{Chen:2023wpj,
    author = "Chen, Anson and Gray, Rachel and Baker, Tessa",
    title = "{Testing the nature of gravitational wave propagation using dark sirens and galaxy catalogues}",
    eprint = "2309.03833",
    archivePrefix = "arXiv",
    primaryClass = "gr-qc",
    doi = "10.1088/1475-7516/2024/02/035",
    journal = "JCAP",
    volume = "02",
    pages = "035",
    year = "2024"
}

@article{Colangeli:2025bnb,
    author = "Colangeli, Elena and Leyde, Konstantin and Baker, Tessa",
    title = "{A bright future? Prospects for cosmological tests of GR with multimessenger gravitational wave events}",
    eprint = "2501.05560",
    archivePrefix = "arXiv",
    primaryClass = "gr-qc",
    doi = "10.1088/1475-7516/2025/05/078",
    journal = "JCAP",
    volume = "05",
    pages = "078",
    year = "2025"
}

@article{LIGOScientific:2026uyd,
    collaboration = "LIGO Scientific, VIRGO, KAGRA",
    author = {{The LIGO Scientific Collaboration} and {The Virgo Collaboration} and {The KAGRA Collaboration}},
       title = "{GWTC-5.0: Constraints on the Cosmic Expansion Rate and Modified Gravitational-wave Propagation}",
      journal = {arXiv e-prints},
          year = 2026,
         month = may,
           eid = {arXiv:2605.27227},
         pages = {arXiv:2605.27227},
archivePrefix = {arXiv},
        eprint = {2605.27227},
 primaryClass = {astro-ph.CO},
        adsurl = {https://ui.adsabs.harvard.edu/abs/2026arXiv260527227T}
}

@article{Mastrogiovanni:2021wsd,
    author = "Mastrogiovanni, S. and Leyde, K. and Karathanasis, C. and Chassande-Mottin, E. and Steer, D. A. and Gair, J. and Ghosh, A. and Gray, R. and Mukherjee, S. and Rinaldi, S.",
    title = "{On the importance of source population models for gravitational-wave cosmology}",
    eprint = "2103.14663",
    archivePrefix = "arXiv",
    primaryClass = "gr-qc",
    doi = "10.1103/PhysRevD.104.062009",
    journal = "Phys. Rev. D",
    volume = "104",
    number = "6",
    pages = "062009",
    year = "2021"
}

@article{Oguri:2016dgk,
    author = "Oguri, Masamune",
    title = "{Measuring the distance-redshift relation with the cross-correlation of gravitational wave standard sirens and galaxies}",
    eprint = "1603.02356",
    archivePrefix = "arXiv",
    primaryClass = "astro-ph.CO",
    doi = "10.1103/PhysRevD.93.083511",
    journal = "Phys. Rev. D",
    volume = "93",
    number = "8",
    pages = "083511",
    year = "2016"
}

@article{Ferri:2024amc,
    author = "Ferri, Jo{\~a}o and Tashiro, Ian L. and Abramo, L. Raul and Matos, Isabela and Quartin, Miguel and Sturani, Riccardo",
    title = "{A robust cosmic standard ruler from the cross-correlations of galaxies and dark sirens}",
    eprint = "2412.00202",
    archivePrefix = "arXiv",
    primaryClass = "astro-ph.CO",
    reportNumber = "ET-0544A-24",
    doi = "10.1088/1475-7516/2025/04/008",
    journal = "JCAP",
    volume = "04",
    pages = "008",
    year = "2025"
}

@article{SantiagodeMatos:2025iyj,
    author = "Santiago de Matos, Isabela and Dalang, Charles and Baker, Tessa and Abramo, Raul and Ferri, Jo{\~a}o and Quartin, Miguel",
    title = "{First measurement of the Hubble constant from gravitational wave-galaxy cross-correlations}",
      journal = {arXiv e-prints},
         year = 2025,
        month = dec,
          eid = {arXiv:2512.15380},
        pages = {arXiv:2512.15380},
          doi = {10.48550/arXiv.2512.15380},
archivePrefix = {arXiv},
       eprint = {2512.15380},
 primaryClass = {astro-ph.CO},
       adsurl = {https://ui.adsabs.harvard.edu/abs/2025arXiv251215380S}
}

@article{Mukherjee:2020hyn,
    author = "Mukherjee, Suvodip and Wandelt, Benjamin D. and Nissanke, Samaya M. and Silvestri, Alessandra",
    title = "{Accurate precision Cosmology with redshift unknown gravitational wave sources}",
    eprint = "2007.02943",
    archivePrefix = "arXiv",
    primaryClass = "astro-ph.CO",
    doi = "10.1103/PhysRevD.103.043520",
    journal = "Phys. Rev. D",
    volume = "103",
    number = "4",
    pages = "043520",
    year = "2021"
}

@article{Ezquiaga:2022zkx,
    author = "Ezquiaga, Jose Mar{\'\i}a and Holz, Daniel E.",
    title = "{Spectral Sirens: Cosmology from the Full Mass Distribution of Compact Binaries}",
    eprint = "2202.08240",
    archivePrefix = "arXiv",
    primaryClass = "astro-ph.CO",
    doi = "10.1103/PhysRevLett.129.061102",
    journal = "Phys. Rev. Lett.",
    volume = "129",
    number = "6",
    pages = "061102",
    year = "2022"
}

@article{Mukherjee:2018ebj,
    author = "Mukherjee, Suvodip and Wandelt, Benjamin D.",
        title = "{Beyond the classical distance-redshift test: cross-correlating redshift-free standard candles and sirens with redshift surveys}",
      journal = {arXiv e-prints},
          year = 2018,
         month = aug,
           eid = {arXiv:1808.06615},
         pages = {arXiv:1808.06615},
           doi = {10.48550/arXiv.1808.06615},
archivePrefix = {arXiv},
        eprint = {1808.06615},
 primaryClass = {astro-ph.CO},
        adsurl = {https://ui.adsabs.harvard.edu/abs/2018arXiv180806615M}
}

@article{Mancarella:2021ecn,
    author = "Mancarella, Michele and Genoud-Prachex, Edwin and Maggiore, Michele",
    title = "{Cosmology and modified gravitational wave propagation from binary black hole population models}",
    eprint = "2112.05728",
    archivePrefix = "arXiv",
    primaryClass = "gr-qc",
    doi = "10.1103/PhysRevD.105.064030",
    journal = "Phys. Rev. D",
    volume = "105",
    number = "6",
    pages = "064030",
    year = "2022"
}

@article{Gray:2019ksv,
    author = "Gray, Rachel and others",
    title = "{Cosmological inference using gravitational wave standard sirens: A mock data analysis}",
    eprint = "1908.06050",
    archivePrefix = "arXiv",
    primaryClass = "gr-qc",
    reportNumber = "LIGO-P1900017",
    doi = "10.1103/PhysRevD.101.122001",
    journal = "Phys. Rev. D",
    volume = "101",
    number = "12",
    pages = "122001",
    year = "2020"
}

@article{DelPozzo:2011vcw,
    author = "Del Pozzo, Walter",
    title = "{Inference of the cosmological parameters from gravitational waves: application to second generation interferometers}",
    eprint = "1108.1317",
    archivePrefix = "arXiv",
    primaryClass = "astro-ph.CO",
    doi = "10.1103/PhysRevD.86.043011",
    journal = "Phys. Rev. D",
    volume = "86",
    pages = "043011",
    year = "2012"
}

@article{Tagliazucchi:2026gxn,
    author = "Tagliazucchi, Matteo and Moresco, Michele and Borghi, Nicola and Ciapetti, Chiara",
    title = "{Mind the peak: Improving cosmological constraints from GWTC-4.0 spectral sirens using semiparametric mass models}",
    eprint = "2601.03347",
    archivePrefix = "arXiv",
    primaryClass = "astro-ph.CO",
    doi = "10.1051/0004-6361/202558756",
    journal = "Astron. Astrophys.",
    volume = "709",
    pages = "A197",
    year = "2026"
}

@article{LIGOScientific:2026ctl,
    collaboration = "LIGO Scientific, VIRGO, KAGRA",
    author = {{The LIGO Scientific Collaboration} and {The Virgo Collaboration} and {The KAGRA Collaboration}},
 title = "{GWTC-5.0: Population Properties of Merging Compact Binaries}",
      journal = {arXiv e-prints},
          year = 2026,
         month = may,
           eid = {arXiv:2605.27226},
         pages = {arXiv:2605.27226},
archivePrefix = {arXiv},
        eprint = {2605.27226},
 primaryClass = {astro-ph.HE},
        adsurl = {https://ui.adsabs.harvard.edu/abs/2026arXiv260527226T}
}

@article{Finke:2021aom,
    author = "Finke, Andreas and Foffa, Stefano and Iacovelli, Francesco and Maggiore, Michele and Mancarella, Michele",
    title = "{Cosmology with LIGO/Virgo dark sirens: Hubble parameter and modified gravitational wave propagation}",
    eprint = "2101.12660",
    archivePrefix = "arXiv",
    primaryClass = "astro-ph.CO",
    doi = "10.1088/1475-7516/2021/08/026",
    journal = "JCAP",
    volume = "08",
    pages = "026",
    year = "2021"
}

@article{Belgacem:2018lbp,
    author = "Belgacem, Enis and Dirian, Yves and Foffa, Stefano and Maggiore, Michele",
    title = "{Modified gravitational-wave propagation and standard sirens}",
    eprint = "1805.08731",
    archivePrefix = "arXiv",
    primaryClass = "gr-qc",
    doi = "10.1103/PhysRevD.98.023510",
    journal = "Phys. Rev. D",
    volume = "98",
    number = "2",
    pages = "023510",
    year = "2018"
}

@article{Belgacem:2017ihm,
    author = "Belgacem, Enis and Dirian, Yves and Foffa, Stefano and Maggiore, Michele",
    title = "{Gravitational-wave luminosity distance in modified gravity theories}",
    eprint = "1712.08108",
    archivePrefix = "arXiv",
    primaryClass = "astro-ph.CO",
    doi = "10.1103/PhysRevD.97.104066",
    journal = "Phys. Rev. D",
    volume = "97",
    number = "10",
    pages = "104066",
    year = "2018"
}

@article{Bellini:2014fua,
    author = "Bellini, Emilio and Sawicki, Ignacy",
    title = "{Maximal freedom at minimum cost: linear large-scale structure in general modifications of gravity}",
    eprint = "1404.3713",
    archivePrefix = "arXiv",
    primaryClass = "astro-ph.CO",
    doi = "10.1088/1475-7516/2014/07/050",
    journal = "JCAP",
    volume = "07",
    pages = "050",
    year = "2014"
}

@article{Linder:2016wqw,
    author = "Linder, Eric V.",
    title = "{Challenges in connecting modified gravity theory and observations}",
    eprint = "1607.03113",
    archivePrefix = "arXiv",
    primaryClass = "astro-ph.CO",
    doi = "10.1103/PhysRevD.95.023518",
    journal = "Phys. Rev. D",
    volume = "95",
    number = "2",
    pages = "023518",
    year = "2017"
}

@article{Leyde:2022orh,
    author = "Leyde, Konstantin and Mastrogiovanni, Simone and Steer, Dani{\`e}le A. and Chassande-Mottin, Eric and Karathanasis, Christos",
    title = "{Current and future constraints on cosmology and modified gravitational wave friction from binary black holes}",
    eprint = "2202.00025",
    archivePrefix = "arXiv",
    primaryClass = "gr-qc",
    doi = "10.1088/1475-7516/2022/09/012",
    journal = "JCAP",
    volume = "09",
    pages = "012",
    year = "2022"
}

@article{Bertheas:2026odj,
    author = "Bertheas, Tom and Gennari, Vasco and Steer, Dani{\`e}le and Tamanini, Nicola",
    title = "{Spectral sirens cosmology from binary black holes populations with sharper mass features}",
    eprint = "2603.06792",
    archivePrefix = "arXiv",
    adsurl="https://arxiv.org/abs/2603.06792",
    primaryClass = "gr-qc",
    pages = "arXiv:2603.06792",
    month = "3",
    year = "2026",
    journal = {arXiv e-prints}
}

@article{Gennari:2026dfy,
    author = "Gennari, Vasco and Bertheas, Tom and Tamanini, Nicola",
    title = "{Emergent structure in the binary black hole mass distribution and implications for population-based cosmology}",
    eprint = "2604.14290",
    adsurl = "https://arxiv.org/abs/2604.14290",
    archivePrefix = "arXiv",
    primaryClass = "gr-qc",
    month = "4",
    year = "2026",
    journal = {arXiv e-prints}
}

@article{Ezquiaga:2021ler,
    author = "Ezquiaga, Jose Maria and Hu, Wayne and Lagos, Macarena and Lin, Meng-Xiang",
    title = "{Gravitational wave propagation beyond general relativity: waveform distortions and echoes}",
    eprint = "2108.10872",
    archivePrefix = "arXiv",
    primaryClass = "astro-ph.CO",
    doi = "10.1088/1475-7516/2021/11/048",
    journal = "JCAP",
    volume = "11",
    number = "11",
    pages = "048",
    year = "2021"
}

@article{Ezquiaga:2018btd,
    author = "Ezquiaga, Jose Mar{\'\i}a and Zumalac{\'a}rregui, Miguel",
    title = "{Dark Energy in light of Multi-Messenger Gravitational-Wave astronomy}",
    eprint = "1807.09241",
    archivePrefix = "arXiv",
    primaryClass = "astro-ph.CO",
    reportNumber = "IFT-UAM-CSIC-18-83, CERN-TH-2018-172",
    doi = "10.3389/fspas.2018.00044",
    journal = "Front. Astron. Space Sci.",
    volume = "5",
    pages = "44",
    year = "2018"
}

@article{LISACosmologyWorkingGroup:2019mwx,
    author = "Belgacem, Enis and others",
    collaboration = "LISA Cosmology Working Group",
    title = "{Testing modified gravity at cosmological distances with LISA standard sirens}",
    eprint = "1906.01593",
    archivePrefix = "arXiv",
    primaryClass = "astro-ph.CO",
    reportNumber = "LISA CosWG-19-01; IFT-UAM-CSIC-19-79, LISA CosWG-19-01",
    doi = "10.1088/1475-7516/2019/07/024",
    journal = "JCAP",
    volume = "07",
    pages = "024",
    year = "2019"
}

@article{COSMOPYRO,
    author  = "Leyde, Konstantin and Elena Colangeli",
    title   = "{CosmoPyro: Gradients for Gravitational-Wave Cosmology}",
    year    = "2026",
    eid = "2608.18281",
    pages = "2608.18281",
    archivePrefix = {arXiv},
    journal = {arXiv e-prints},
    eprint = {2608.18281},
    adsurl = "https://arxiv.org/abs/2608.18281",
    primaryClass={astro-ph.CO},
    month = "8"
    }

@article{Romano:2022jeh,
    author = "Romano, Antonio Enea",
    title = "{Effective speed of gravitational waves}",
    eprint = "2211.05760",
    archivePrefix = "arXiv",
    primaryClass = "gr-qc",
    doi = "10.1016/j.physletb.2024.138572",
    journal = "Phys. Lett. B",
    volume = "851",
    pages = "138572",
    year = "2024"
}

@article{Mukherjee:2020mha,
    author = "Mukherjee, Suvodip and Wandelt, Benjamin D. and Silk, Joseph",
    title = "{Testing the general theory of relativity using gravitational wave propagation from dark standard sirens}",
    eprint = "2012.15316",
    archivePrefix = "arXiv",
    primaryClass = "astro-ph.CO",
    doi = "10.1093/mnras/stab001",
    journal = "Mon. Not. Roy. Astron. Soc.",
    volume = "502",
    number = "1",
    pages = "1136--1144",
    year = "2021"
}

@article{Amendola:2017ovw,
    author = "Amendola, Luca and Sawicki, Ignacy and Kunz, Martin and Saltas, Ippocratis D.",
    title = "{Direct detection of gravitational waves can measure the time variation of the Planck mass}",
    eprint = "1712.08623",
    archivePrefix = "arXiv",
    primaryClass = "astro-ph.CO",
    doi = "10.1088/1475-7516/2018/08/030",
    journal = "JCAP",
    volume = "08",
    pages = "030",
    year = "2018"
}

@article{Saltas:2014dha,
    author = "Saltas, Ippocratis D. and Sawicki, Ignacy and Amendola, Luca and Kunz, Martin",
    title = "{Anisotropic Stress as a Signature of Nonstandard Propagation of Gravitational Waves}",
    eprint = "1406.7139",
    archivePrefix = "arXiv",
    primaryClass = "astro-ph.CO",
    doi = "10.1103/PhysRevLett.113.191101",
    journal = "Phys. Rev. Lett.",
    volume = "113",
    number = "19",
    pages = "191101",
    year = "2014"
}

@article{Leyde:2024tov,
    author = "Leyde, Konstantin and Baker, Tessa and Enzi, Wolfgang",
    title = "{Cosmic cartography: Bayesian reconstruction of the galaxy density informed by large-scale structure}",
    eprint = "2409.20531",
    archivePrefix = "arXiv",
    primaryClass = "astro-ph.CO",
    doi = "10.1088/1475-7516/2024/12/013",
    journal = "JCAP",
    volume = "12",
    pages = "013",
    year = "2024"
}

@article{Heinzel:2025ogf,
    author = "Heinzel, Jack and Vitale, Salvatore",
    title = "{When (not) to trust Monte Carlo approximations for hierarchical Bayesian inference}",
    eid = "arXiv:2509.07221",
    pages= "arXiv:2509.07221",
    archivePrefix = {arXiv},
    eprint = {2509.07221},
    adsurl = "https://arxiv.org/abs/2509.07221",
    primaryClass = "astro-ph.HE",
    month = "9",
    year = "2025",
    journal = {arXiv e-prints}
    
}

@article{hoffman2014no,
author = {Homan, Matthew D. and Gelman, Andrew},
title = {The No-U-turn sampler: adaptively setting path lengths in Hamiltonian Monte Carlo},
year = {2014},
issue_date = {January 2014},
publisher = {JMLR.org},
volume = {15},
number = {1},
issn = {1532-4435},
journal = {J. Mach. Learn. Res.},
month = jan,
pages = {1593–1623},
numpages = {31},
adsurl="https://dl.acm.org/doi/10.5555/2627435.2638586"}

@article{LIGOScientific:2025slb,
    author = "Abac, A. G. and others",
    collaboration = "LIGO Scientific, VIRGO, KAGRA",
    title = "{GWTC-4.0: Updating the Gravitational-Wave Transient Catalog with Observations from the First Part of the Fourth LIGO-Virgo-KAGRA Observing Run}",
    eprint = "2508.18082",
    archivePrefix = "arXiv",
    eid="arXiv:2508.18082",
    pages="arXiv:2508.18082",
    primaryClass = "gr-qc",
    reportNumber = "LIGO-P2400386",
    doi = "10.3847/2041-8213/ae2c74",
    journal = "Astrophys. J. Lett.",
    volume = "1004",
    number = "2",
    year = "2026"
}

@article{LIGOScientific:2026wfs,
    author = {{The LIGO Scientific Collaboration} and {The Virgo Collaboration} and {The KAGRA Collaboration}},
    collaboration = "LIGO Scientific, VIRGO, KAGRA",
title = "{GWTC-5.0: Observations from the Second Part of the Fourth LIGO-Virgo-KAGRA Observing Run and Updates to the Gravitational-Wave Transient Catalog}",
      journal = {arXiv e-prints},
          year = 2026,
         month = may,
           eid = {arXiv:2605.27225},
         pages = {arXiv:2605.27225},
archivePrefix = {arXiv},
        eprint = {2605.27225},
 primaryClass = {gr-qc},
        adsurl = {https://ui.adsabs.harvard.edu/abs/2026arXiv260527225T}
}

@article{ET:2025xjr,
    author = "Abac, Adrian and others",
    collaboration = "ET",
    title = "{The Science of the Einstein Telescope}",
    eprint = "2503.12263",
    archivePrefix = "arXiv",
    primaryClass = "gr-qc",
    reportNumber = "ET-0036C-25",
    doi = "10.1088/1475-7516/2026/03/081",
    journal = "JCAP",
    volume = "03",
    pages = "081",
    year = "2026"
}

@article{Cutler:1994ys,
    author = "Cutler, Curt and Flanagan, Eanna E.",
    title = "{Gravitational waves from merging compact binaries: How accurately can one extract the binary's parameters from the inspiral wave form?}",
    eprint = "gr-qc/9402014",
    archivePrefix = "arXiv",
    reportNumber = "GRP-369",
    doi = "10.1103/PhysRevD.49.2658",
    journal = "Phys. Rev. D",
    volume = "49",
    pages = "2658--2697",
    year = "1994"
}

@article{Maggiore:2024cwf,
    author = "Maggiore, Michele and Iacovelli, Francesco and Belgacem, Enis and Mancarella, Michele and Muttoni, Niccol{\`o}",
    title = "{Comparison of global networks of third-generation gravitational-wave detectors}",
    eprint = "2411.05754",
    archivePrefix = "arXiv",
    primaryClass = "gr-qc",
    doi = "10.1088/1361-6382/ae110b",
    journal = "Class. Quant. Grav.",
    volume = "42",
    number = "21",
    pages = "215004",
    year = "2025"
}

@article{Gupta:2023lga,
    author = "Gupta, Ish and others",
    title = "{Characterizing gravitational wave detector networks: from A$^\sharp$ to cosmic explorer}",
    eprint = "2307.10421",
    archivePrefix = "arXiv",
    primaryClass = "gr-qc",
    reportNumber = "CE Document No. P2300019, CE Document No. P2300019-v2",
    doi = "10.1088/1361-6382/ad7b99",
    journal = "Class. Quant. Grav.",
    volume = "41",
    number = "24",
    pages = "245001",
    year = "2024"
}

@article{Nicholl:2024ttg,
    author = "Nicholl, Matt and Andreoni, Igor",
    title = "{Electromagnetic follow-up of gravitational waves: review and lessons learned}",
    eprint = "2410.18274",
    archivePrefix = "arXiv",
    primaryClass = "astro-ph.HE",
    doi = "10.1098/rsta.2024.0126",
    journal = "Phil. Trans. Roy. Soc. Lond. A",
    volume = "383",
    number = "2294",
    pages = "20240126",
    year = "2025"
}

@article{Keinan:2024gai,
    author = "Keinan, Ido and Arcavi, Iair",
    title = "{The Potential of Coordinated Gravitational-wave Follow-up for Improving Kilonova Detection Prospects: Lessons from GW190425}",
    eprint = "2405.17558",
    archivePrefix = "arXiv",
    primaryClass = "astro-ph.HE",
    doi = "10.3847/1538-4357/adcba5",
    journal = "Astrophys. J.",
    volume = "985",
    number = "1",
    pages = "142",
    year = "2025"
}

@article{Steeghs:2021wcr,
    author = "Steeghs, D. and others",
    title = "{The Gravitational-wave Optical Transient Observer (GOTO): prototype performance and prospects for transient science}",
    eprint = "2110.05539",
    archivePrefix = "arXiv",
    primaryClass = "astro-ph.IM",
    doi = "10.1093/mnras/stac013",
    journal = "Mon. Not. Roy. Astron. Soc.",
    volume = "511",
    number = "2",
    pages = "2405--2422",
    year = "2022"
}

@article{Riess:2021jrx,
    author = "Riess, Adam G. and others",
    title = "{A Comprehensive Measurement of the Local Value of the Hubble Constant with 1 km s$^{−1}$ Mpc$^{−1}$ Uncertainty from the Hubble Space Telescope and the SH0ES Team}",
    eprint = "2112.04510",
    archivePrefix = "arXiv",
    primaryClass = "astro-ph.CO",
    doi = "10.3847/2041-8213/ac5c5b",
    journal = "Astrophys. J. Lett.",
    volume = "934",
    number = "1",
    pages = "L7",
    year = "2022"
}

@article{Duane:1987de,
    author = "Duane, S. and Kennedy, A. D. and Pendleton, B. J. and Roweth, D.",
    title = "{Hybrid Monte Carlo}",
    doi = "10.1016/0370-2693(87)91197-X",
    journal = "Phys. Lett. B",
    volume = "195",
    pages = "216--222",
    year = "1987"
}

@article{Flanagan:1997sx,
    author = "Flanagan, Eanna E. and Hughes, Scott A.",
    title = "{Measuring gravitational waves from binary black hole coalescences: 1. Signal-to-noise for inspiral, merger, and ringdown}",
    eprint = "gr-qc/9701039",
    archivePrefix = "arXiv",
    reportNumber = "GRP-456",
    doi = "10.1103/PhysRevD.57.4535",
    journal = "Phys. Rev. D",
    volume = "57",
    pages = "4535--4565",
    year = "1998"
}

@article{KAGRA:2013rdx,
    author = "Abbott, B. P. and others",
    collaboration = "KAGRA, LIGO Scientific, Virgo",
    title = "{Prospects for observing and localizing gravitational-wave transients with Advanced LIGO, Advanced Virgo and KAGRA}",
    eprint = "1304.0670",
    archivePrefix = "arXiv",
    primaryClass = "gr-qc",
    reportNumber = "LIGO-P1200087, VIR-0288A-12, JGW-P1808427",
    doi = "10.1007/s41114-020-00026-9",
    journal = "Living Rev. Rel.",
    volume = "19",
    pages = "1",
    year = "2016"
}

@article{Dominik:2014yma,
    author = "Dominik, Michal and Berti, Emanuele and O'Shaughnessy, Richard and Mandel, Ilya and Belczynski, Krzysztof and Fryer, Christopher and Holz, Daniel E. and Bulik, Tomasz and Pannarale, Francesco",
    title = "{Double Compact Objects III: Gravitational Wave Detection Rates}",
    eprint = "1405.7016",
    archivePrefix = "arXiv",
    primaryClass = "astro-ph.HE",
    doi = "10.1088/0004-637X/806/2/263",
    journal = "Astrophys. J.",
    volume = "806",
    number = "2",
    pages = "263",
    year = "2015"
}

@article{Belgacem:2019tbw,
    author = "Belgacem, Enis and Dirian, Yves and Foffa, Stefano and Howell, Eric J. and Maggiore, Michele and Regimbau, Tania",
    title = "{Cosmology and dark energy from joint gravitational wave-GRB observations}",
    eprint = "1907.01487",
    archivePrefix = "arXiv",
    primaryClass = "astro-ph.CO",
    doi = "10.1088/1475-7516/2019/08/015",
    journal = "JCAP",
    volume = "08",
    pages = "015",
    year = "2019"
}

@article{Nishizawa:2019rra,
    author = "Nishizawa, Atsushi and Arai, Shun",
    title = "{Generalized framework for testing gravity with gravitational-wave propagation. III. Future prospect}",
    eprint = "1901.08249",
    archivePrefix = "arXiv",
    primaryClass = "gr-qc",
    doi = "10.1103/PhysRevD.99.104038",
    journal = "Phys. Rev. D",
    volume = "99",
    number = "10",
    pages = "104038",
    year = "2019"
}

@article{DAgostino:2019hvh,
    author = "D'Agostino, Rocco and Nunes, Rafael C.",
    title = "{Probing observational bounds on scalar-tensor theories from standard sirens}",
    eprint = "1907.05516",
    archivePrefix = "arXiv",
    primaryClass = "gr-qc",
    doi = "10.1103/PhysRevD.100.044041",
    journal = "Phys. Rev. D",
    volume = "100",
    number = "4",
    pages = "044041",
    year = "2019"
}

@article{Chen:2026htz,
    author = "Chen, Anson and Zhang, Jun",
    title = "{Forecasting constraints on cosmology and modified gravitational-wave propagation by combining strongly lensed gravitational waves and galaxy surveys}",
    eprint = "2601.21820",
    archivePrefix = "arXiv",
    primaryClass = "astro-ph.CO",
    doi = "10.1103/prcx-d577",
    journal = "Phys. Rev. D",
    volume = "113",
    number = "12",
    pages = "123535",
    year = "2026"
}

@article{Chen:2026owi,
    author = "Chen, Anson and Zhang, Jun",
        title = "{Searching for Extra Dimensions with Gravitational Waves: Dark-Siren Constraints from GWTC-4}",
      journal = {arXiv e-prints},
          year = 2026,
         month = jun,
           eid = {arXiv:2606.14549},
         pages = {arXiv:2606.14549},
           doi = {10.48550/arXiv.2606.14549},
archivePrefix = {arXiv},
        eprint = {2606.14549},
 primaryClass = {gr-qc},
        adsurl = {https://ui.adsabs.harvard.edu/abs/2026arXiv260614549C}
}

@article{Borghi:2023opd,
    author = "Borghi, Nicola and Mancarella, Michele and Moresco, Michele and Tagliazucchi, Matteo and Iacovelli, Francesco and Cimatti, Andrea and Maggiore, Michele",
    title = "{Cosmology and Astrophysics with Standard Sirens and Galaxy Catalogs in View of Future Gravitational Wave Observations}",
    eprint = "2312.05302",
    archivePrefix = "arXiv",
    primaryClass = "astro-ph.CO",
    doi = "10.3847/1538-4357/ad20eb",
    journal = "Astrophys. J.",
    volume = "964",
    number = "2",
    pages = "191",
    year = "2024"
}

@article{Tagliazucchi:2025ofb,
    author = "Tagliazucchi, Matteo and Moresco, Michele and Borghi, Nicola and Fiebig, Manfred",
        title = "{Accelerating the standard siren method: Improved constraints on modified gravitational-wave propagation with future data}",
      journal = {\aap},
          year = 2025,
         month = oct,
        volume = {702},
           eid = {A244},
         pages = {A244},
           doi = {10.1051/0004-6361/202554827},
archivePrefix = {arXiv},
        eprint = {2504.02034},
 primaryClass = {astro-ph.CO},
        adsurl = {https://ui.adsabs.harvard.edu/abs/2025A&A...702A.244T}
}

@article{Gray:2021sew,
    author = "Gray, Rachel and Messenger, Chris and Veitch, John",
    title = "{A pixelated approach to galaxy catalogue incompleteness: improving the dark siren measurement of the Hubble constant}",
    eprint = "2111.04629",
    archivePrefix = "arXiv",
    primaryClass = "astro-ph.CO",
    doi = "10.1093/mnras/stac366",
    journal = "Mon. Not. Roy. Astron. Soc.",
    volume = "512",
    number = "1",
    pages = "1127--1140",
    year = "2022"
}

@article{Kiziltan:2013oja,
    author = {Kiziltan, B{\"u}lent and Kottas, Athanasios and De Yoreo, Maria and Thorsett, Stephen E.},
    title = "{The Neutron Star Mass Distribution}",
    eprint = "1309.6635",
    archivePrefix = "arXiv",
    primaryClass = "astro-ph.SR",
    doi = "10.1088/0004-637X/778/1/66",
    journal = "Astrophys. J.",
    volume = "778",
    pages = "66",
    year = "2013"
}

@article{LIGOScientific:2020aai,
    author = "Abbott, B. P. and others",
    collaboration = "LIGO Scientific, Virgo",
    title = "{GW190425: Observation of a Compact Binary Coalescence with Total Mass $\sim 3.4 M_{\odot}$}",
    eprint = "2001.01761",
    archivePrefix = "arXiv",
    primaryClass = "astro-ph.HE",
    reportNumber = "LIGO-P190425",
    doi = "10.3847/2041-8213/ab75f5",
    journal = "Astrophys. J. Lett.",
    volume = "892",
    number = "1",
    pages = "L3",
    year = "2020"
}

@article{DESI:2025zgx,
    author = "Abdul Karim, M. and others",
    collaboration = "DESI",
    title = "{DESI DR2 results. II. Measurements of baryon acoustic oscillations and cosmological constraints}",
    eprint = "2503.14738",
    archivePrefix = "arXiv",
    primaryClass = "astro-ph.CO",
    reportNumber = "FERMILAB-PUB-25-0169-PPD",
    doi = "10.1103/tr6y-kpc6",
    journal = "Phys. Rev. D",
    volume = "112",
    number = "8",
    pages = "083515",
    year = "2025"
}

@article{Brout:2022vxf,
    author = "Brout, Dillon and others",
    title = "{The Pantheon+ Analysis: Cosmological Constraints}",
    eprint = "2202.04077",
    archivePrefix = "arXiv",
    primaryClass = "astro-ph.CO",
    doi = "10.3847/1538-4357/ac8e04",
    journal = "Astrophys. J.",
    volume = "938",
    number = "2",
    pages = "110",
    year = "2022"
}

@article{Dupletsa:2026uqs,
    author = "Dupletsa, Ulyana and others",
        title = "{Radio sirens: inferring $H_0$ with binary black holes and neutral hydrogen in the era of the Einstein Telescope and the SKA Observatory}",
      journal = {arXiv e-prints},
          year = 2026,
         month = may,
           eid = {arXiv:2605.12606},
         pages = {arXiv:2605.12606},
           doi = {10.48550/arXiv.2605.12606},
archivePrefix = {arXiv},
        eprint = {2605.12606},
 primaryClass = {astro-ph.CO},
        adsurl = {https://ui.adsabs.harvard.edu/abs/2026arXiv260512606D}
}

@article{Zazzera:2025ord,
    author = "Zazzera, Stefano and Fonseca, Jos{\'e} and Baker, Tessa and Clarkson, Chris",
        title = "{Exploring future synergies for large-scale structure between gravitational waves and radio sources}",
        journal = {\mnras},
         year = 2026,
        month = mar,
       volume = {547},
       number = {1},
          eid = {stag307},
        pages = {stag307},
          doi = {10.1093/mnras/stag307},
archivePrefix = {arXiv},
       eprint = {2505.15645},
 primaryClass = {astro-ph.CO},
       adsurl = {https://ui.adsabs.harvard.edu/abs/2026MNRAS.547ag307Z}
}

@article{Horndeski:1974wa,
    author = "Horndeski, Gregory Walter",
        title = "{Second-order scalar-tensor field equations in a four-dimensional space}",
      journal = {International Journal of Theoretical Physics},
         year = 1974,
        month = sep,
       volume = {10},
       number = {6},
        pages = {363-384},
          doi = {10.1007/BF01807638},
       adsurl = {https://ui.adsabs.harvard.edu/abs/1974IJTP...10..363H}
}

@article{Pogosian:2021mcs,
    author = "Pogosian, Levon and Raveri, Marco and Koyama, Kazuya and Martinelli, Matteo and Silvestri, Alessandra and Zhao, Gong-Bo and Li, Jian and Peirone, Simone and Zucca, Alex",
        title = "{Imprints of cosmological tensions in reconstructed gravity}",
      journal = {Nature Astronomy},
         year = 2022,
        month = dec,
       volume = {6},
        pages = {1484-1490},
          doi = {10.1038/s41550-022-01808-7},
archivePrefix = {arXiv},
       eprint = {2107.12992},
 primaryClass = {astro-ph.CO},
       adsurl = {https://ui.adsabs.harvard.edu/abs/2022NatAs...6.1484P}
}

@article{Raveri:2021dbu,
    author = "Raveri, Marco and Pogosian, Levon and Martinelli, Matteo and Koyama, Kazuya and Silvestri, Alessandra and Zhao, Gong-Bo",
  title = "{Principal reconstructed modes of dark energy and gravity}",
      journal = {\jcap},
         year = 2023,
        month = feb,
       volume = {2023},
       number = {2},
          eid = {061},
        pages = {061},
          doi = {10.1088/1475-7516/2023/02/061},
archivePrefix = {arXiv},
       eprint = {2107.12990},
 primaryClass = {astro-ph.CO},
       adsurl = {https://ui.adsabs.harvard.edu/abs/2023JCAP...02..061R}
}

@article{Finn:1992xs,
    author = "Finn, Lee Samuel and Chernoff, David F.",
    title = "{Observing binary inspiral in gravitational radiation: One interferometer}",
    eprint = "gr-qc/9301003",
    archivePrefix = "arXiv",
    reportNumber = "PRINT-93-0138 (NORTHWESTERN)",
    doi = "10.1103/PhysRevD.47.2198",
    journal = "Phys. Rev. D",
    volume = "47",
    pages = "2198--2219",
    year = "1993"
}
